\newcommand{\caproman}[1]{\uppercase\expandafter{\romannumeral#1}}

\newcommand{\BGmail}{bgottsch\,@\,fas.harvard.edu}

\documentclass[titlepage]{article}

\title{\bf Techniques of Proton Radiotherapy:\\Transport Theory}
\author{B. Gottschalk\thanks{Harvard University Laboratory for Particle Physics and Cosmology, 18 Hammond St., Cambridge, MA 02138, USA, \BGmail}}

\usepackage{amsmath}
\usepackage{latexsym}
\usepackage{graphicx}

\begin{document}

\maketitle

\begin{abstract}
These are lecture notes, in three main parts: phase space diagrams, Fermi-Eyges theory, and the application of the theory to transverse beam spreading in homogeneous matter.

Phase space diagrams serve to visualize correlations between position and direction in a particle beam. They do not in themselves yield quantitative formulas, but they facilitate geometric reasoning which makes such formulas more plausible. We introduce phase space via a model beam line where multiple Coulomb scattering (MCS) and drift (motion along the beam line) happen separately (usually they are combined). We discuss the effect of a pure scatterer, and of a pure drift, on the phase space distribution. We introduce the beam ellipse and the area enclosed by it, the emittance. We discuss how emittance changes in a drift and in a thin scatterer. Finally, we discuss phase space diagrams of a more realistic beam line.

In a brief detour, we review various mathematical topics, justify the Gaussian approximation by comparing it to the full theory of MCS, and briefly discuss the concept `scattering power' which is needed in Fermi-Eyges theory.

When the phase space distribution happens to be Gaussian it can be computed by means of Fermi-Eyges theory. The most general case is a Gaussian incident beam traversing one or more homogeneous slabs of varying materials. All the resulting phase space distributions will be Gaussian until the beam stops or encounters a beam limiting device or a transverse heterogeneity. We discuss Eyges' original theory for an ideal beam entering a single slab, and its later generalization to a finite (but still Gaussian) beam entering a stack of slabs. We revisit the beam ellipse, showing it to be an iso-density contour in phase space. We use the theory to confirm the emittance change in a drift and in a thin scatterer. We discuss equivalent sources and finally, we compute the fraction of the beam contained by the beam ellipse defined in a more general way.

Preston and Koehler (PK), without using Fermi-Eyges theory, studied transverse spreading of particle beams stopping in matter. We recast their derivations in Fermi-Eyges language. First, we show that their basic equation for the rms beam size at any measuring plane, which they obtained by direct physical reasoning, is equivalent to a basic equation of Fermi-Eyges theory. We recover their result that the transverse beam size at end-of-range is proportional to range, and their further result that (beam size)/(maximum size) versus (depth)/(maximum depth) is a universal function (which they express in closed form), independent of incident energy and stopping material. We generalize both results to arbitrary ion species, and we review the experimental data for protons and ions.

\end{abstract}

\clearpage
\tableofcontents

\clearpage
\section{Introduction}
From time to time we teach a one week intensive course `Techniques of Proton Radiotherapy' to persons involved in the design, upgrade or maintenance of proton radiotherapy facilities, or otherwise interested. An undergraduate degree in Physics or Engineering is required, but familiarity with particle physics is not. The course is equal parts basic physics, engineering (using the physics for useful calculations) and experimental devices and techniques. Lectures, a draft textbook, and some useful Fortran software and executables are available for free download \cite{BGcourse,BGware}.

These notes deal with deterministic transport theory, the topic (above all others) which time does not permit us to cover fully in the classroom. By this point in the course we have studied energy loss and multiple Coulomb scattering (MCS) of charged particles, and it is time to see what happens when those processes occur simultaneously with drift (motion along the beam axis).

\begin{figure}[h]
\centering\includegraphics[width=4.87in,height=3.5in]{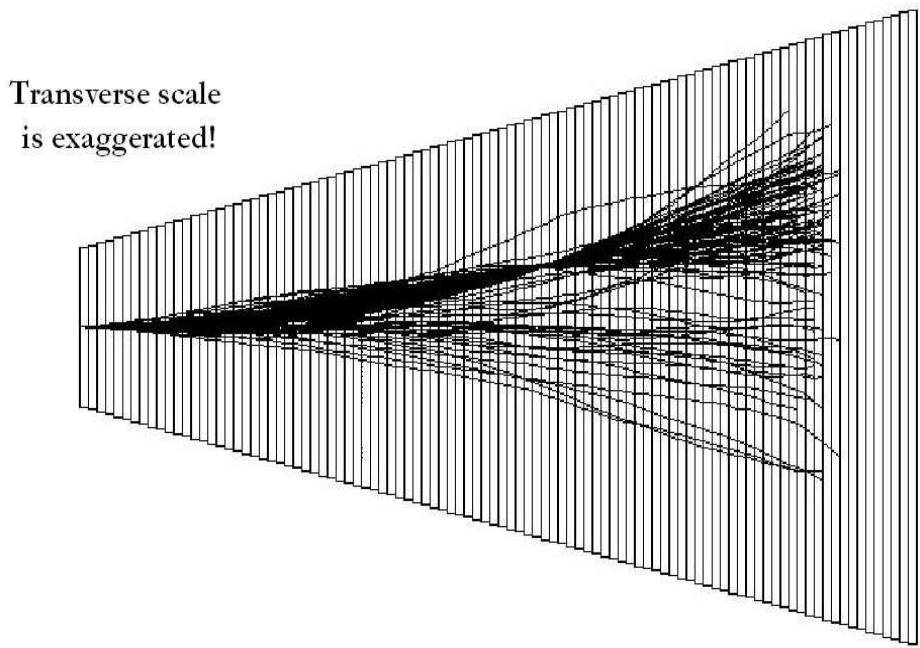} %
\caption{Top view of protons stopping in a water tank, with transverse scale exaggerated. Fermi-Eyges theory predicts  the spatial and angular distributions of the full beam at any depth, and of the protons passing through the off-axis slit.\label{fig:TransportWater}}
\end{figure}

Figure\,\ref{fig:TransportWater} shows a proton beam stopping in a water tank. The transverse scale is exaggerated (actual proton angles with respect to the beam are only a few degrees). The tracks were simulated by a simple Monte Carlo (MC) program. The first thirty were plotted without conditions. Thereafter, only those passing through a virtual off-axis slit were drawn. We'll sketch how the MC works, in order to contrast it with the main body of these notes.

The MC divides the water into layers or slabs. Starting at the first slab, a proton is `transported' as follows. Its incoming position and direction are projected to the  midpoint, and its energy reduced by an amount appropriate to a half-slab. $x$ and $y$ deflections are chosen at random from a probability distribution that obeys the laws of MCS for water, that slab thickness, and that proton energy. The new directions are used to project the track to the downstream face of the slab, and another half-slab of energy is deducted. Given incoming position, direction and energy we now have outgoing position, direction and energy. We repeat for the next slab, and so on until the proton stops (energy reduced to zero). Then the whole thing is repeated for the next proton.

Although real MCs such as Geant4 are more complicated, their basic principle is the same. Final quantities such as the rms spread of the beam at some depth, or the fluence at some point, are found by scoring the MC results. Accuracy is limited by statistics, that is, by how many proton `histories' we have run. Thus the MC method is non-deterministic: its results change, consistent with statistics, from run to run.\footnote{~To be sure, if the random number seed and everything else in the computer run is unchanged, we'll get exactly the same answer every time. Nevertheless, the statistical uncertainty is still there, but less obvious.} Unfortunately, Monte Carlo is the best way presently known to solve problems where the geometry through which the protons propagate is complicated.

If, however, the geometry is simple (as in Figure\,\ref{fig:TransportWater}) a deterministic and much faster method is available: Fermi-Eyges (FE) theory, the main subject of these notes. Because of the restricted geometry, FE by itself solves relatively few practical problems. Among those are single scattering beam lines, the front end of double scattering beam lines, beam spreading in matter, and the emittance increase in a cyclotron energy degrader. Nevertheless it is worth studying because it is a building block in many other problems, such as pencil beam dose algorithms and finding the most likely path of a proton through a degrader. FE gives us a feeling for how a beam behaves in simple cases, before we tackle more complicated ones.

\section{Phase Space Diagrams}
\begin{figure}[p]
\centering\includegraphics[width=4.55in,height=3.5in]{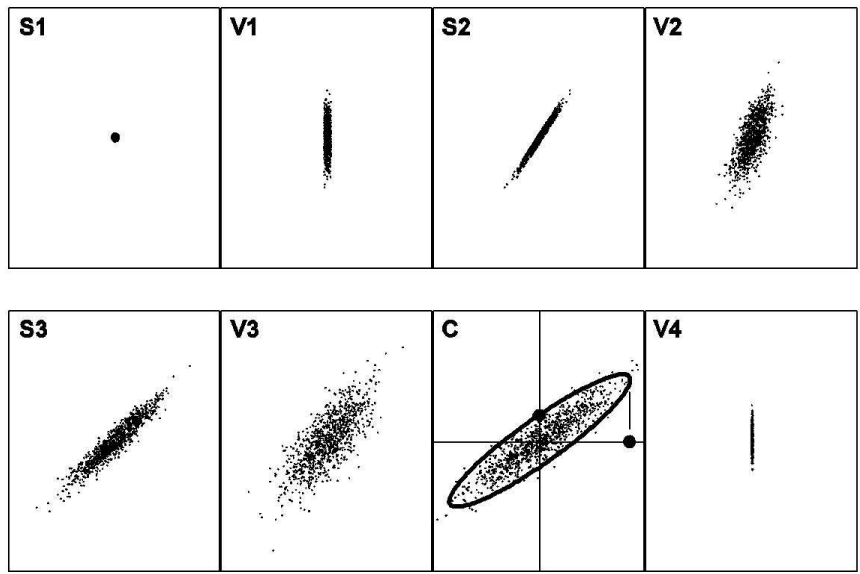} 
\caption{Phase space evolution of the model beam (distributions at the {\em entrance} to each labeled element). For visual clarity the ellipse is drawn at $2.5\,\sigma$ rather than the conventional $1\,\sigma$; the standard ellipse would only contain about 390 of the 1000 phase space points.\label{fig:ModelPS}}
\end{figure}

As Figure\;\ref{fig:TransportWater} shows, relationships between proton positions and directions are complicated even in a simple situation. Phase space diagrams give us a systematic way of visualizing them geometrically, even if we have to bring in other methods to get quantitative results. In the phase space picture each proton propagates along the beam direction $z$ and we examine its phase space variables
\[x\quad x'\quad y\quad y'\quad E\quad\]
at various `depths' or values of $z$. $x,y$ are transverse positions and $x',y'$ are slopes, which represent the proton's direction. In small angle approximation, which is always valid for radiotherapy protons, the slopes equal the projected angles $\theta_x,\;\theta_y$. (When there is no possibility of confusion, we'll drop the subscript and refer to $\theta_x$ simply as $\theta$. In other words, we'll use $x'$, $\theta_x$ and $\theta$ more or less interchangeably.) $E$ is the kinetic energy.\footnote{~In medical physics $E$ is commonly used for kinetic energy, though $T$ is sometimes used \cite{icru49}. In general physics, $E$ usually means total energy \cite{Livingston1954} as in $E=mc^2$. We reserve $T$ for scattering power.}  A `phase space diagram' is a scatter plot of $x'$ vs. $x$, or $y'$ vs. $y$, at a given value of $z$. For a sample, look ahead at Figure\;\ref{fig:ModelPS}. Very soon, you will understand what is going on here and why such diagrams are useful.

In the phase space picture $E$ is a sort of `ghost' variable, always present but not actually shown. It begins at the incident energy (say 160\;Mev) and decreases as the protons progress down the beam line. When we start doing quantitative calculations we will need to keep track of it (that is, `transport' it) because---for instance---the strength of a given scatterer located at some $z$ will depend on the energy of the protons at that $z$. In simple cases, $E$ decreases as a function of $z$ in more or less the same way for all protons, so each snapshot at a given $z$ can be labeled with some $E$. In more complicated cases (for instance, the patient) protons with significantly different energies may reach the same point $x,y,z$ because of heterogeneities and multiple Coulomb scattering. Such situations are beyond the scope of these notes.

In some problems it may be more convenient to use a variable other than $E$ as the `longitudinal' variable. For instance, in problems involving magnets, the momentum $pc$ (MeV) is convenient, whereas in dosimetry we may use the residual range in water (cm). Always, the longitudinal variable is some measure of the proton's speed.

Phase space diagrams and the concept of the beam ellipse are much used in connection with particle accelerator and magnetic beam line design \cite{carey}. Their explicit use in papers on scattered beams is relatively recent, though even the early development of Fermi-Eyges theory implies a phase-space picture in the background.

In oriented materials, scattering may be very different in $x$ and $y$. We only consider homogeneous scatterers, where the scattering in $x$ and $y$ is the same, so it is sufficient to draw a diagram for one or the other. We'll use $x,\,\theta\,(=\theta_x=x')$. That works until the point in the beam line where a beam limiting device (collimator) or something else breaks the symmetry between $x$ and $y$.

\begin{figure}[p]
\centering\includegraphics[width=4.87in,height=3.5in]{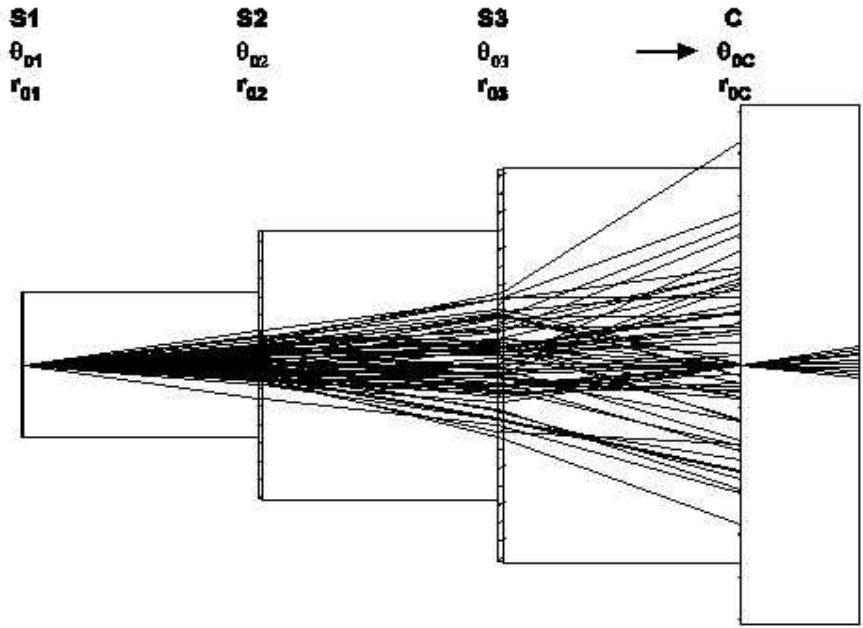} 
\caption{Model beam line for phase space analysis: three scatterers separated by voids or drifts, a collimator with a central slit, and a final drift.\label{fig:ModelBeam}}
\end{figure}

\subsection{Model Beam Line}
Let's  introduce phase space with a beam line of no practical use. It is a `separated' beam line consisting of three thin scatters (MCS and energy loss only, no drift) separated by voids or `drifts' (drift only, no MCS or energy loss), followed by an ideal collimator having a small slit on axis, followed by a final drift. In any real slab of matter MCS, energy loss and drift occur simultaneously. Eventually we'll take that into account. In the beginning it's easier to separate them.

Figure\,\ref{fig:ModelBeam} shows proton rays in the model beam line with, as always, the transverse dimension greatly magnified. Figure\,\ref{fig:ModelPS} is a preview of the phase space diagrams corresponding to those rays. Each frame shows phase space going {\em into} that beam line element; thus S3 labels the phase space diagram obtaining at the {\em entrance} to the third scatterer. The next two sections explain these phase space diagrams.

\subsection{Effect of a Scatterer}

\begin{figure}[p]
\centering\includegraphics[width=6.0in,height=2.33in]{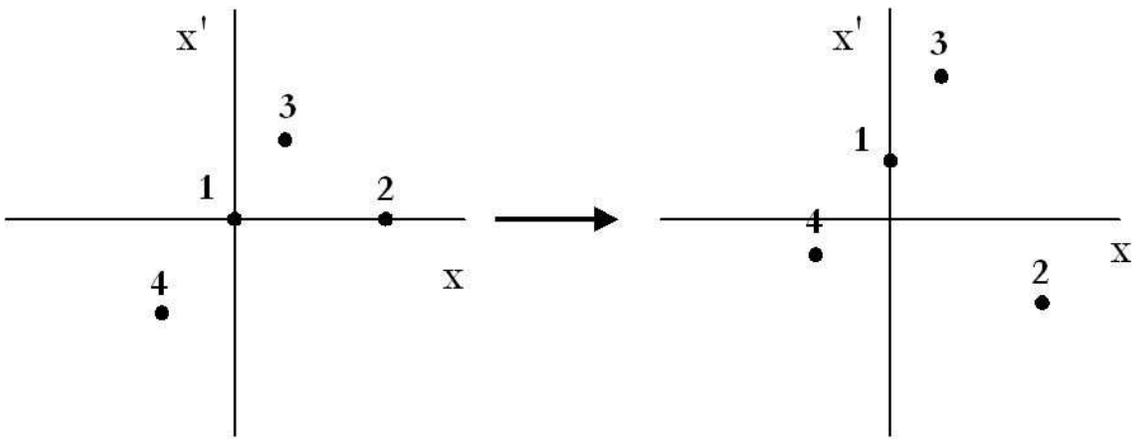} 
\caption{Phase space diagram of four protons acted on by a scatterer.\label{fig:Scatterer}}
\end{figure}

In a scatterer, points representing single protons receive random vertical kicks. The size of the kick has a Gaussian distribution with a width that depends on the scatterer material and thickness, and on the proton energy at that $z$. The distribution of kicks is independent of $x$ for a uniform scatterer. The kicks may be positive, negative, small (more likely) or large. Figure\,\ref{fig:Scatterer} illustrates all that. No point is untouched, except the lucky proton that does not scatter. However, because there is no drift in a thin scatterer, the horizontal position of each point is unaffected. 

\subsection{Effect of a Drift}
\begin{figure}[p]
\centering\includegraphics[width=6.0in,height=3.88in]{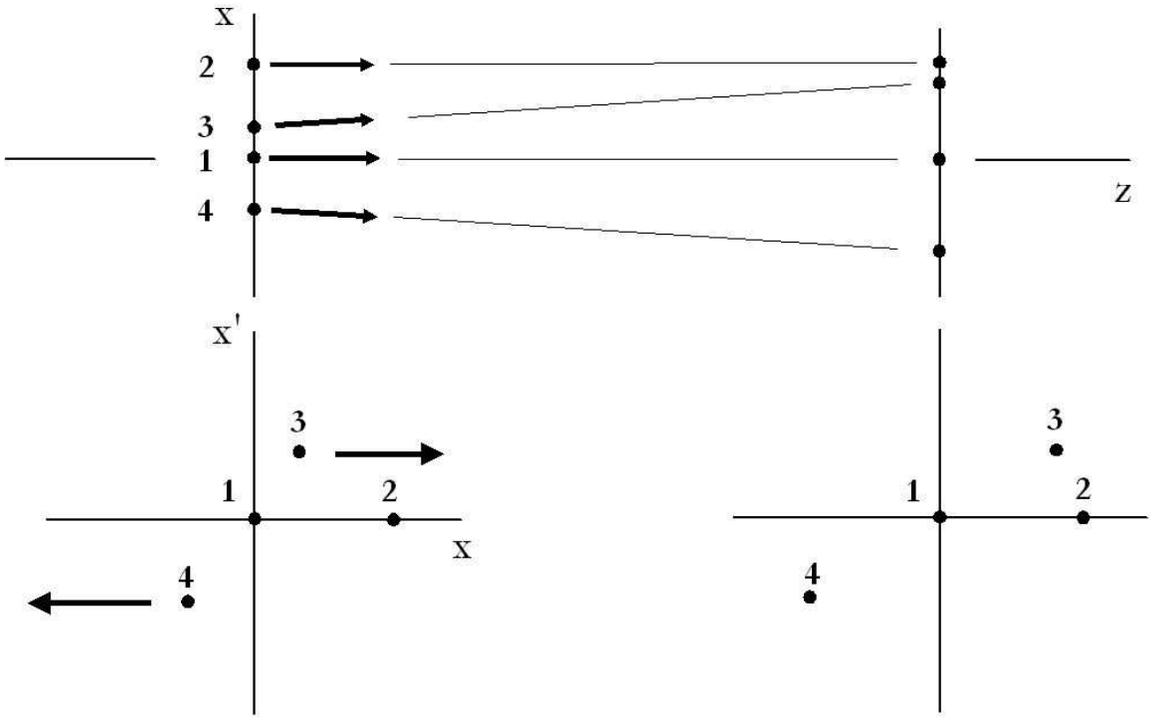} 
\caption{Phase space diagram of four protons acted on by a drift.\label{fig:Drift}}
\end{figure}

In a drift, points in the upper half move to the right in proportion to their distance from the $x$ axis whereas points in the lower half move to the left. In a pure drift (no scattering) the vertical position of each point is unaffected. The net result is a shear (not a rotation!) of the phase space distribution. Points on the $x$ axis stay where they are. Figure\,\ref{fig:Drift} illustrates all this. To help we have drawn the proton trajectories ($x$ vs. $z$) in the top part of the figure.

\subsection{The Beam Ellipse}
A subset of points in the phase space diagram can be surrounded by an ellipse. The entire behavior of the beam is encapsulated in the behavior of that ellipse, as we'll see.

In scattering/drift beam lines (even when the two occur simultaneously) the ellipse has a rigorous quantitative meaning. It is an iso-density contour in phase space, as we'll show. That contrasts with accelerator theory (magnetic beam transport) where the beam ellipse is merely a convenience. It contains a defined fraction of the beam, but is not directly related to the phase space density.

Figure\,\ref{fig:ModelPS} frame C (phase space at collimator entrance) shows a typical beam ellipse. The area enclosed by the beam ellipse is an important property of the beam known as its emittance. In the Appendix we show that the area enclosed by a tilted centered ellipse equals $\pi \,x'_\mathrm{int}\,x_\mathrm{max}$ (or, by symmetry, $\pi\,x'_\mathrm{max}\,x_\mathrm{int}$).\footnote{~That reduces to the more familiar ($\pi ab$) if the ellipse is erect.} The area enclosed by a degenerate ellipse (point or line segment), and therefore the emittance of the beam so described, is zero. That means a beam can have a finite size (extent in $x$) or divergence (extent in $x'$) or both, and still have zero emittance. Zero emittance means that position and angle are perfectly correlated, not that either is necessarily zero. 


\begin{figure}[p]
\centering\includegraphics[width=3.90in,height=3.5in]{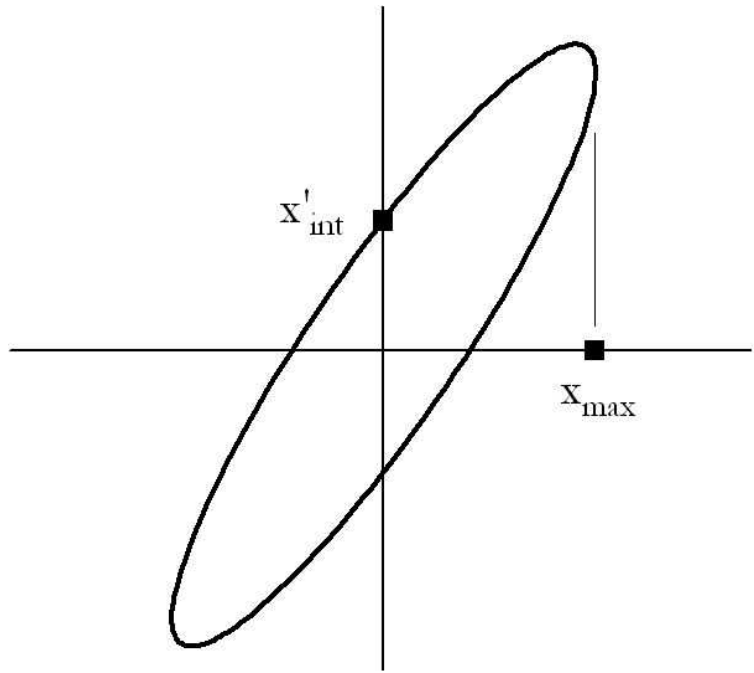} 
\caption{The beam ellipse, which encloses an area $\pi\;x'_\mathrm{int}\;x_\mathrm{max}=
\pi\;x'_\mathrm{max}\;x_\mathrm{int}$.\label{fig:EllipseArea}}
\end{figure}
\begin{figure}[p]
\centering\includegraphics[width=4.90in,height=3.5in]{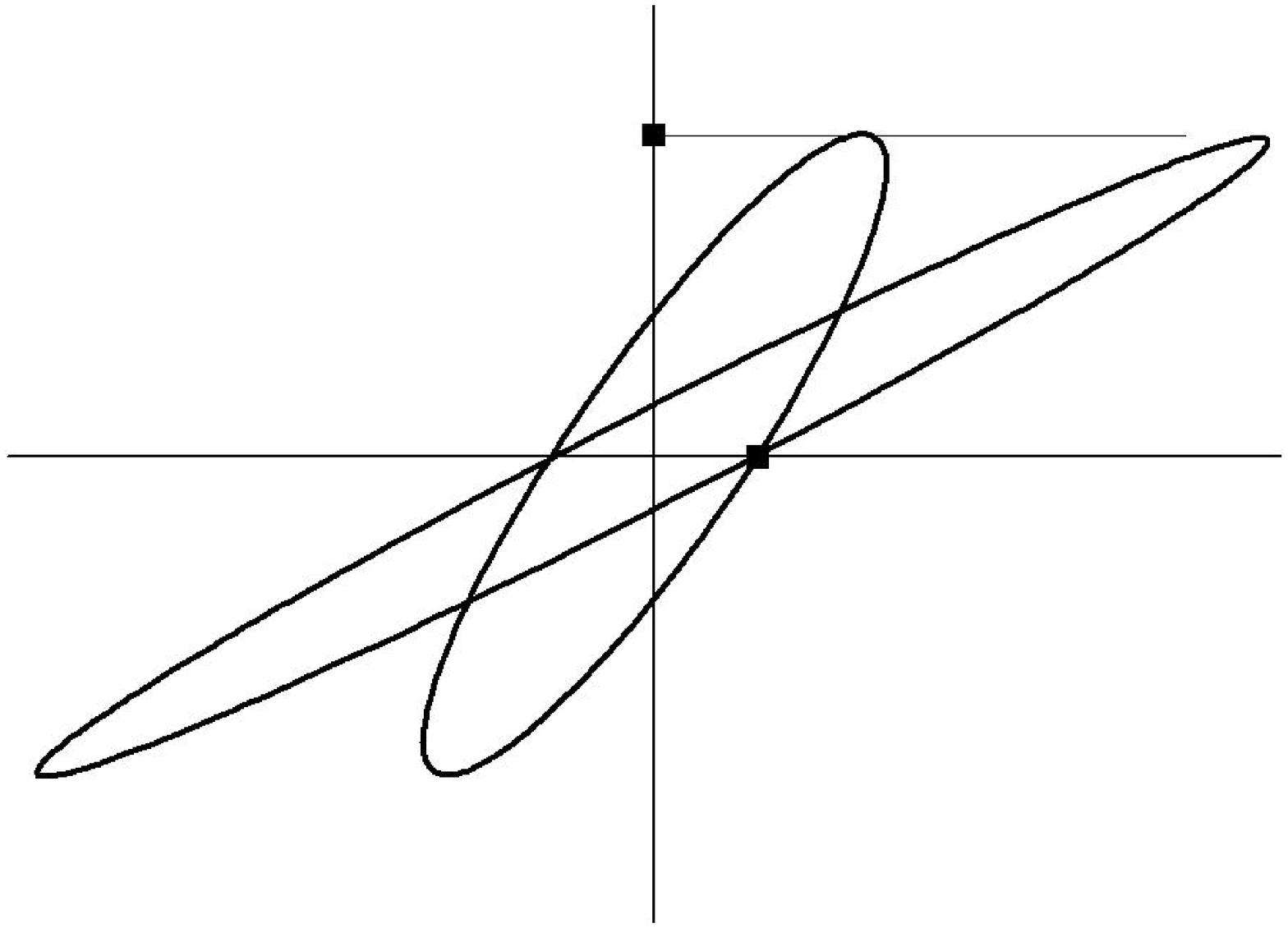} 
\caption{Liouville's Theorem: the area enclosed by the beam ellipse (the beam emittance) is conserved in a drift.\label{fig:Liouville}}
\end{figure}

\subsection{Phase Space Diagrams for the Model Beam Line}
We are now ready to understand the phase space diagrams in Figure\,\ref{fig:ModelPS}.\footnote{~In early dealings with phase space diagrams you will be tempted to think of them as cross sectional pictures such as might be obtained by putting a piece of film in the beam. That would in essence be a scatter plot of $y$ vs. $x$. That is not what they are! They are (more abstract) plots of $x'$ vs. $x$.} The first frame shows the ideal beam entering the first scatterer. The beam has neither spatial extent nor divergence, so all of the protons fall on a single point at the origin of phase space. (For better visual effect, we have drawn a small blob instead.)

Leaving the scatterer into the first void, the beam has acquired a distribution of angles but still no size. Entering the second scatterer phase space has sheared. The beam now has both size and angular spread but $x'$ and $x$ are perfectly correlated and the phase space area (had the incident beam truly been a point) is still zero.

Leaving the second scatterer, however, the phase space distribution will have an area no matter how perfect the incident beam. The exact correlation between angle and position is destroyed. `Angular confusion' has been introduced.\footnote{~The term `angular confusion' was introduced by Andy Koehler in discussing protons \cite{urie}. The fact that standard Fermi-Eyges notation for the same quantity is $\theta_C$ \cite{icru35} may be fortuitous.}

You can follow the rest by yourself: a shear, more scatter, another shear. We have drawn an ellipse around the beam entering the collimator.\footnote{~For visual effect we have drawn the ellipse much larger than would be usual, at $2.5\times\sigma$ of the underlying Gaussian distributions, whereas the standard ellipse of which we will speak later would be at $1\times\sigma$.}

Finally, a few of the protons pass through a narrow slit in the collimator. The rest stop. Protons emerging have no spatial spread left (except the slit width) but they have an angular spread. The rms value of that spread $\theta_C$ is the quantitative definition of the angular confusion of the beam. Note that we have solved---by Monte Carlo simulation---a problem that might at first seem difficult: If an ideal incident beam makes its way down a sequence of scatterers and drifts, what is the angular spread of protons that happen to pass through a slit at the end? When we study Fermi-Eyges theory, we'll learn how to compute $\theta_C$ another way, without simulation.

Soon we'll also learn why we're interested in $\theta_C$\,. When we compute the transverse penumbra of a scattered beam (the sharpness of the shadow cast by a collimator and therefore, the sharpness of the dose distribution) $\theta_C$ turns out to be the key and it, in turn, depends on the emittance as should be obvious from frame C of Figure\,\ref{fig:ModelPS}. No emittance, no $\theta_C$. 

\subsection{Emittance Change in a Drift}
Figure\,\ref{fig:Liouville} shows a beam ellipse before and after a drift. $x_\mathrm{int}$ does not change because points on the $x$ axis do not move in a drift. $x'_\mathrm{max}$ does not change because there is no scattering. From our area formula we therefore conclude that  the area enclosed by the beam ellipse, that is, the emittance, is conserved in a drift. This is a special case of Liouville's Theorem. 

\subsection{Emittance Change in a Scatterer}
Consider Figure\,\ref{fig:ModelPS} panel V1. Scattering by itself does not increase emittance. If the incident beam were truly perfect, the phase space distribution in V1 (degenerate ellipse) would still have no area. Emittance first comes about unavoidably in panel V2 just after the second scatterer. Two factors are needed for an emittance increase: the beam must have some size, or spread in $x$, and the scatterer must have some finite strength, that is, cause some random up or down kicks in $\theta$. Indeed, we will show later that the increase (in quadrature) of the emittance in a thin scatterer equals exactly $\pi$ times the product of the rms size of the beam at the scatterer times the rms angular deflection caused by the scatterer. A given scatterer will increase the emittance less, if it is further upstream where the beam is smaller. That becomes an important consideration in beamline design.

\begin{figure}[p]
\centering\includegraphics[width=4.55in,height=3.5in]{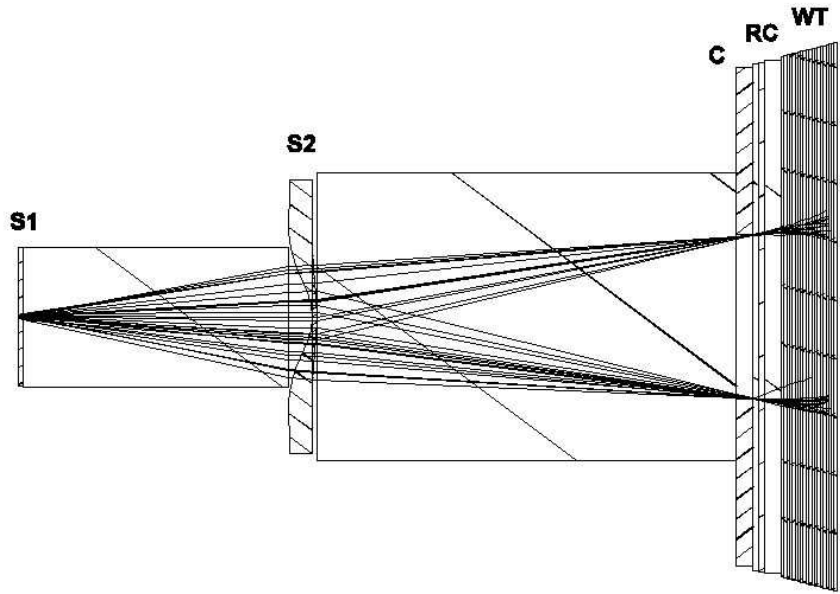} 
\caption{Double scattered beam line with selected trajectories of protons that graze the downstream edges of the patient collimator. S1 represents one step of a range modulator.\label{fig:DSbeam}}
\end{figure}
\begin{figure}[p]
\centering\includegraphics[width=4.55in,height=3.5in]{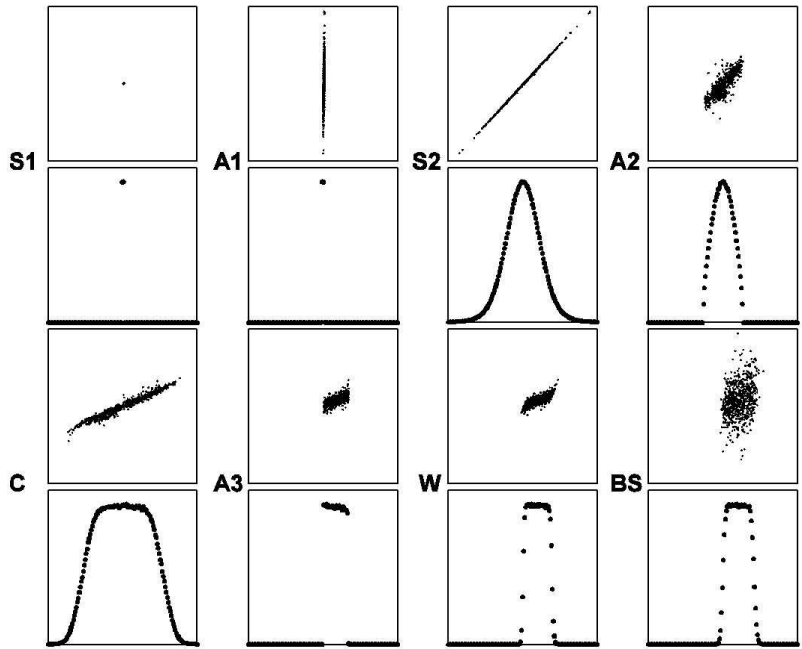} 
\caption{Phase space diagrams for the double scattered beam with $x$ projections interleaved. (n.b. The setup used for these differs in two points from Figure\,\ref{fig:DSbeam}. There is a collimator surrounding S2, and the `patient collimator' is a half-beam block with one edge on axis rather than the symmetric one of Figure\,\ref{fig:DSbeam}.)\label{fig:DSphase}}
\end{figure}

\subsection{Phase Space for a More Realistic Beam Line}
Figure\,\ref{fig:DSbeam} shows a modern double scattered passive beam line.\footnote{~The first scatterer represents an upstream range modulator stopped at one of its $M$ thickness steps. To solve the full problem, we'll eventually need to solve it for each step separately, then sum over $M$ using appropriate weights.} We have drawn selected Monte Carlo generated rays, ones that pass through a virtual slit near each collimator edge. These justify our interest in $\theta_C$ since the unsharpness of the collimator's shadow clearly depends on the rms cone angle of protons grazing the collimator. It also clearly depends on how far downstream the water tank is, showing why air gap is so important in proton radiotherapy. The dose edge is further degraded by scattering in the water tank, showing why depth in the patient is yet another important factor.

Figure\,\ref{fig:DSphase} shows the same beam in phase space, interleaved with projections onto the $x$ axis to show fluence distributions in $x$. The first three panels are as before. The fluence entering the second scatterer is Gaussian. However, leaving S2 (entering the second air gap) two new things have happened. A collimator (not shown in Figure\,\ref{fig:DSbeam}) around S2 cuts off phase space at some radius, producing sharp edges. Also, S2 is stronger, and therefore the vertical kicks are greater, near the beam axis. The ensuing greater shear as protons drift to the patient collimator C is, by design, just enough to flatten the fluence distribution.

The patient collimator chops off protons on the left and right creating sharp edges at A3. Those edges shear in the ensuing drift so their projections entering the water tank smear out, causing penumbra which depends not only on the angular confusion (vertical extent of the phase space distribution) at C but also on the length of the drift (air gap). Multiple scattering and drift in the water tank blur the edges still further.

\subsection{Summary}
\begin{quote}
The phase space coordinates of a proton at depth $z$ are $x,\;x',\;y,\;y'$ and $E$.

A phase space diagram is a scatter plot of $x'$ vs, $x$. Each point represents one proton. The diagram is not a cross section of the beam!

In a thin scatterer each point acquires a random $\pm$ vertical kick, whose rms size depends on the strength of the scatterer. Its $x$ position is unchanged.

In a pure drift, the phase space distribution shears, points above the $x$ axis moving right in proportion to their distance from the axis, points below moving left. Points on the $x$ axis do not move.

A phase space distribution can be characterized by an ellipse which is an isodensity contour (to be shown).

The beam emittance is the area enclosed by the ellipse and equals $\pi \,x'_\mathrm{int}\,x_\mathrm{max}=\pi\,x'_\mathrm{max}\,x_\mathrm{int}$

Emittance is conserved in a drift, and increases (in quadrature) by $\pi\;x_{rms}\;\theta_0$ in a thin scatterer (to be shown). Thus a given scatterer has greater effect, the larger the incident beam.

The evolution of the beam ellipse (three parameters) can be computed by Fermi-Eyges theory (to be shown). That amounts to complete knowledge of the beam at every $z$ provided the requirements of FE theory are met.

\end{quote}

\section{Miscellaneous Topics}
This section gets a few issues out of the way before we begin Fermi-Eyges theory proper.

\subsection{Review of Gaussians}
The standard 1D Gaussian probability density $P$ and its normalization, mean and variance are
\begin{equation}\label{eqn:PG}
P(u)\;=\;\frac{1}{\sqrt{2\pi}\;\sigma}\;e^{\displaystyle{-\;\frac{1}{2}}\Bigl(\frac{u-u_0}{\sigma}\Bigr)^2}
\end{equation}
\begin{eqnarray}
\int_{-\infty}^\infty P(u)du&=&1\label{eqn:PGnorm}\\
\int_{-\infty}^\infty uP(u)du&=&<u>\;=\;u_0\label{eqn:PGmean}\\
\int_{-\infty}^\infty u^2P(u)du&=&<u^2>\;=\;\sigma^2\label{eqn:PGvariance}
\end{eqnarray}
$P(u)$ stands for a probability {\em density} whose dimensionality is indicated by the number of arguments. Thus $P(x,\theta)\,dx\,d\theta$ is a joint probability, $P(x|\,\theta)\,dx$ is a conditional probability and $NP(x,\theta)$ is a number density in phase space given $N$ incident protons.

The 2D `cylindrical' Gaussian probability density is 
\begin{equation}\label{eqn:PGcyl}
P(r,\phi)\;=\;\frac{1}{2\pi\;r_0^2}\;e^{\displaystyle{-\;\frac{1}{2}}\Bigl(\frac{r}{r_0}\Bigr)^2}
\end{equation}
which is also normalized
\begin{equation}\label{eqn:PGcylnorm}
\int_0^{2\pi}\int_0^\infty P(r,\phi)\,r\,dr\,d\phi\;=\;1
\end{equation}
However
\begin{equation}\label{eqn:PGcylvariance}
\sigma_r^2\;\equiv\;\int_0^{2\pi}\int_0^\infty r^2\,P(r,\phi)\,r\,dr\,d\phi\;=\;2\,r_0^2
\end{equation}
and Eq.\,\ref{eqn:PGcyl} is often written
\begin{equation}\label{eqn:sigr}
P(r,\phi)\;=\;\frac{1}{\pi\;\sigma_r^2}\;e^{\displaystyle{-\;}\Bigl(\frac{r}{\sigma_r}\Bigr)^2}
\end{equation}
When reading a paper, consider whether the authors use Eq.\,\ref{eqn:PGcyl} or Eq.\,\ref{eqn:sigr} because of the $\sqrt{2}$ difference in the Gaussian width parameter ($r_0$ or $\sigma_r$). 

\subsection{The Gaussian Approximation to MCS}
Angular and fluence distributions in Fermi-Eyges theory are Gaussian: FE alone can never predict a non-Gaussian distribution of anything. That flows from the basic assumptions (infinite uniform slabs) and the fact that MCS is very nearly Gaussian. How good is that approximation?

\begin{figure}[h]
\centering\includegraphics[width=4.58in,height=3.5in]{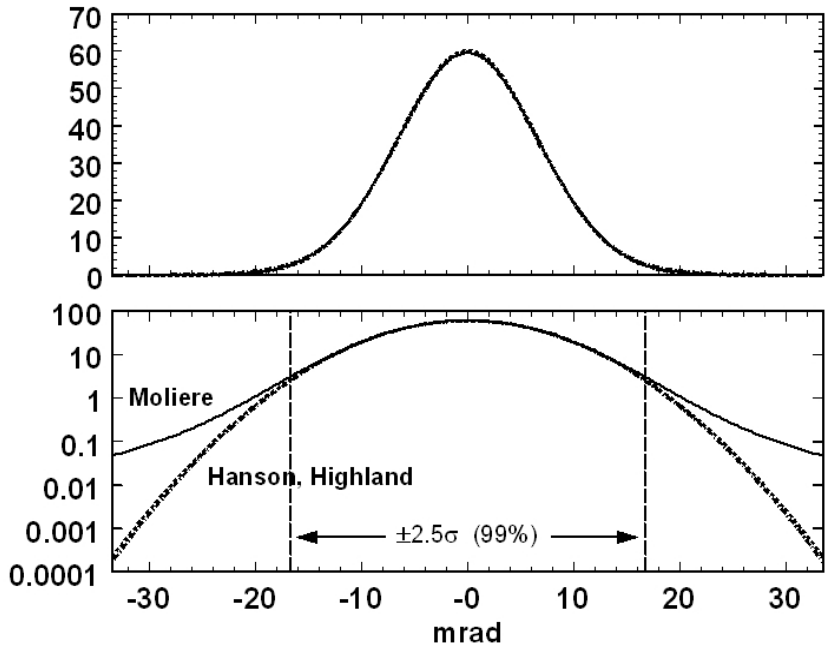} 
\caption{Projected angle distributions for 158.6 MeV protons traversing 1\,cm of water. On a linear plot the
Moli\`ere/Fano distribution is indistinguishable from Gaussians using the Hanson or Highland $\theta_0$. On a log plot, the correct distribution  peels away at $2.5\sigma$, and is more than $100\times$ higher at $5\sigma$.\label{fig:GaussApprox}}
\end{figure}

Figure\,\ref{fig:GaussApprox} shows the outgoing projected angle distribution of 158.6\,MeV protons traversing 1\,cm of water, according to three variants of MCS theory.\footnote{~For details, please see \cite{mcsbg}.} `Highland' is a Gaussian with a width parameter $\theta_0$ given by Highland's formula, a simple and quite accurate parametrization of the full theory. `Hanson' is also a Gaussian, but its width parameter is obtained by computing the full theory, then finding the Gaussian that best approximates it. Finally, `Moli\`ere' is the full theory of the angular distribution. It has been compared to experiment and is known to be correct to a few percent \cite{mcsbg}.

Figure\;\ref{fig:GaussApprox} shows that the central part of the angular distribution is indeed Gaussian out to about 2.5$\,\sigma$, and that the three methods agree very well in that region. The $\pm2.5\,\sigma$ region contains about 99\% of the protons according to the normal probability integral \cite{dwight}. If the projected angle distribution is Gaussian, then in small angle approximation the projected spatial distribution on a screen or measuring plane somewhere downstream will also be Gaussian. That justifies a theory that predicts nothing but Gaussians in the very simple cases for which it holds.

Most problems in proton radiotherapy are treated adequately in the Gaussian approximation, with a width parameter given by one or another version of Moli\`ere theory. In the rare case where that is inadequate, the problem is usually not the `single scattering tail' of the Moli\`ere distribution evident in Figure\;\ref{fig:GaussApprox}, but the halo of charged secondaries from non-elastic nuclear reactions that soon appears around any proton beam traversing matter \cite{pedroniPencil}.

\subsection{Relativistic Single Particle Kinematics}
Any introduction to special relativity will provide the basis for the following equations. If $E$ is the kinetic energy of a particle we define its reduced kinetic energy as
\begin{equation}\label{eqn:tau}
\tau\;\equiv\;E/mc^2
\end{equation}
where $mc^2$ is the particle's rest energy. Then $\beta\equiv v/c$ is given by
\begin{equation}\label{eqn:betasq}
\beta^2\;=\;\frac{\tau+2}{(\tau+1)^2}\,\tau
\end{equation}
where $v$ is the particle's speed and $c$ is the speed of light. The kinematic quantity $pv$, common in MCS theory,  is given by
\begin{equation}\label{eqn:pv}
pv\;=\;\frac{\tau+2}{\tau+1}\,E
\end{equation}
where $p$ is the momentum. For completeness, the momentum itself is given by
\begin{equation}\label{eqn:pcsq}
(pc)^2\;=\;(\tau+2)\,mc^2\,E
\end{equation}
which we would need were we dealing with charged particle transport in magnets. These three equations are convenient because their relativistic and nonrelativistic limits are obvious at a glance and because they avoid small differences between large quantities which can occur in relativistic kinematics if one is not careful.

\subsection{Completing the Square}
The technique called `completing the square' allows us to evaluate certain definite integrals. The formula is
\begin{equation}\label{eqn:complete}
au^2+bu+c=a(u-h)^2+k
\end{equation}
where
\begin{equation}\label{eqn:hk}
h=-\frac{b}{2a}\mathrm{\qquad,\qquad}k=c-\frac{b^2}{4a}
\end{equation}

\subsection{Scattering Power}
An important concept in transport theory is the {\em scattering power}
\begin{equation}\label{eqn:T}
T(z)\;\equiv\;\frac{d<\theta_x^2>}{dz}
\end{equation}
$T$, the rate of change with $z$ of the variance of the projected MCS angle, bears a superficial resemblance to stopping power $S\equiv-dE/dz$, the rate of energy loss. However, the analogy is imperfect.

$S$, for which there is one generally accepted theory \cite{icru49}, is a function of the proton's speed or energy and of atomic properties of the stopping material at the point of interest. If we compute $S$ everywhere in a finite slab and integrate, we obtain the correct energy loss in the slab. But if (following Rossi \cite{rossiBook}) we compute $T$ as a similarly `local' function of speed and stopping material and integrate that, the answer is substantially too large, particularly for thin slabs.

That might not seem to be a problem because the generally accepted theory of multiple Coulomb scattering---Moli\`ere theory \cite{moliere1,moliere2,bethe}---gives $\theta_x$ accurately for arbitrary ions, materials, slab thicknesses, and incident energy. But Moli\`ere theory deals only with finite slabs and is therefore not directly suited to transport calculations, deterministic or Monte Carlo, which require a formula for the instantaneous rate of change of $\theta_x$. $T$ is most usefully understood as satisfying that requirement: a differential approximation to Moli\`ere theory, or a function which, when integrated, reproduces the Moli\`ere/Fano/Hanson angle \cite{mcsbg} more or less accurately.

We find that the formula for $T$ must be {\em nonlocal}. The rate of change of $\theta_x$ with respect to $z$ for (say) 50\,MeV protons in water depends to some extent on how much water they have already gone through. There are different ways of expressing that nonlocality \cite{scatPower2010}.

We repeat here, from \cite{scatPower2010}, a few formulas necessary later. Our own `differential Moli\`ere' scattering power is
\begin{equation}\label{eqn:TdM}
T_\mathrm{dM}\;\equiv\;f_\mathrm{dM}(p_1v_1,pv)\times\left(\frac{E_s}{pv}\right)^2\;\frac{1}{X_S}
\end{equation}
where $E_s=15.0$\;Mev. The correction factor, which measures nonlocality by the decrease in $pv$, is
\begin{eqnarray}\label{eqn:fdM}
f_\mathrm{dM}&\equiv&0.5244+0.1975\log_{10}(1-(pv/p_1v_1)^2)+0.2320\log_{10}(pv/\mathrm{MeV})\nonumber\\
&&-\;0.0098\log_{10}(pv/\mathrm{MeV})\log_{10}(1-(pv/p_1v_1)^2)
\end{eqnarray}
$p_1v_1$ is the initial value and $pv$ is the value at the point of interest. The scattering length $X_S$ (cm) is a property of the material at the point of interest given by
\begin{equation}\label{eqn:XS}
\frac{1}{\rho X_S}\;\equiv\;\alpha N r_e^2\,\frac{Z^2}{A}\left\{2\log(33219\,(AZ)^{-1/3})-1\right\}
\end{equation}
where $\rho$ is density, $\alpha$ is the fine structure constant, $N$ is Avogadro's number, $r_e$ is the classical electron radius and $A,Z$ are the atomic weight and atomic number of the scattering element. 
In compounds or mixtures, atoms act independently (`Bragg rule') and the slab is equivalent to very thin sheets of each constituent in the correct proportion. That picture leads directly to
\begin{equation}\label{eqn:XSBragg}
\frac{1}{\rho X_S}\;=\;\sum_iw_i\left(\frac{1}{\rho X_S}\right)_i
\end{equation}
where $w_i$ is the fraction by weight of the $i^\mathrm{th}$ constituent. $X_S$ is a material property very similar to $X_0$, the radiation length. Table \ref{tbl:Overas} compares $X_S$ with $X_0$ for some materials. (Note that it is {\em mass} scattering length and radiation length that are tabulated.)

When integrated, $T_\mathrm{dM}$ reproduces Moli\`ere/Fano/Hanson theory, as well as experimental data at 158.6\,MeV \cite{mcsbg}, to a few percent over a wide range of materials for normalized slab thicknesses $\Delta z/R$ from 0.001 to 1 \cite{scatPower2010}. There is indirect evidence that it also works for mixed slabs \cite{scatPower2010}, but that has not yet been tested experimentally for $T_\mathrm{dM}$ or any other scattering power. 

The nonlocal correction to $T$ seems paradoxical, implying that the interactions of a proton in water at 50 MeV depend upon how much water has been traversed. However, {\em multiple} Coulomb scattering is not a primitive process! It can be viewed as a statement about the statistics of a large cohort of protons each of which has suffered a large number of atomic encounters. Without the non\-local correction factor their rms spread in angle would be exactly proportional to the square root of slab thickness, but it is not, because MCS is not exactly Gaussian. Therefore the degree of approach to `Gaussianity' also plays a role. That is a statistical statement with no implications for the underlying single scattering process which indeed depends only on a proton's speed and impact parameter at the point of interaction, and on the atomic properties of the scattering material.

\section{Fermi-Eyges Theory}

\subsection{History}
Fermi (concerned with the propagation of cosmic rays) considered the joint probability of position and angle, due to MCS, of a single charged particle entering a uniform medium (the atmosphere). He evidently did not consider his derivation worth publishing, since Rossi and Greisen \cite{rossi} state in a footnote
\begin{quote}
`The developments in this article follow closely a lecture given by Professor Fermi at the University of Chicago in the summer of 1940 and include some unpublished results. The writers wish to express their sincere appreciation to Professor Fermi for allowing them to make use of these results.'
\end{quote}
Later, Eyges \cite{eyges} included the effect of energy loss, which Fermi had ignored. Still later, the theory was generalized to cover a non-ideal (but still Gaussian) incident beam, a stack of slabs of different materials, and improved formulas for the scattering power. We can regard this generalized theory as a way of propagating the beam ellipse (hence the phase space distribution) through homogeneous mixed slabs, and we'll refer to it simply as Fermi-Eyges or FE theory, dropping the `generalized'. 

Here are a few comments for the student reading the early papers to reconcile those with our version. Neither Rossi and Greisen \cite{rossi} nor Eyges \cite{eyges} use the term `scattering power'. However, Rossi's book \cite{rossiBook} uses $\theta_s^2\sim$\,rad$^2$/(g/cm$^2$) and clearly recognizes that some formulas for $\theta_s^2$ are better than others, as well as the issues raised by more accurate theories such as Moli\`ere's. The scattering power more or less built into Eyges' derivation (our $T_\mathrm{FR}$) is the worst possible choice \cite{kanematsu08,scatPower2010}. The term `mass scattering power' was introduced by Brahme \cite{brahme} who was also an early user of FE theory in radiotherapy (electrons). The same paper gives a way (not generally used nowadays) of incorporating Moli\`ere theory into the scattering power. 

Early papers alternate between space and projected angle. We use projected angle exclusively, which makes our scattering power half as large. Also in early papers, depth is commonly expressed in units of the radiation length of the material, which is awkward if there is more than one material. For the same reason we normally use scattering power $T\sim$\,(rad$^2$/cm) rather than mass scattering power $T/\rho\sim$\,(rad$^2$/(g/cm$^2$), range $R$ rather than mass range $\rho R$, and so on.

\begin{figure}[h]
\centering\includegraphics[width=4.91in,height=3.5in]{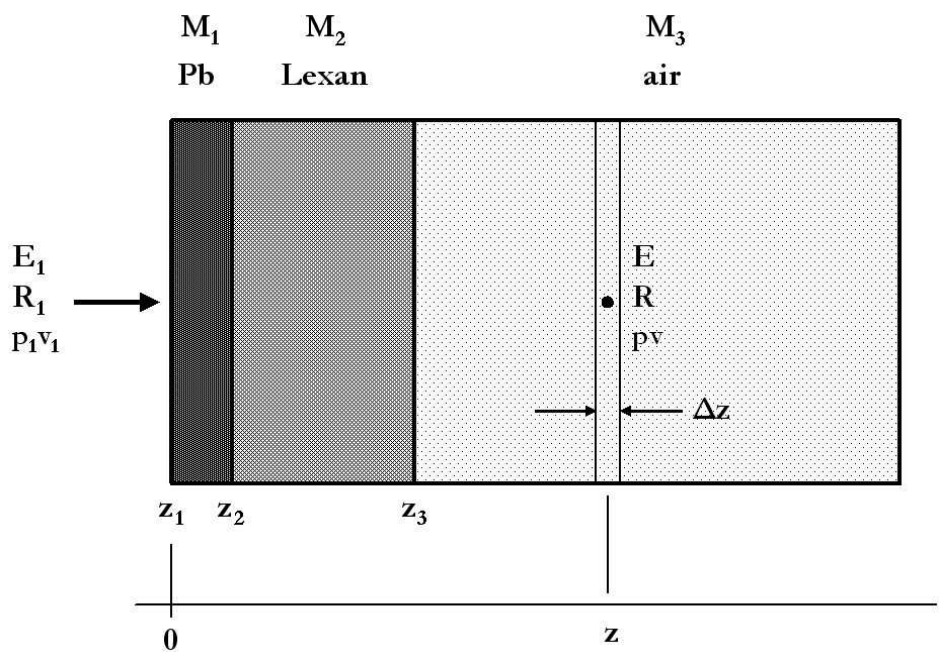} 
\caption{Stack of three homogeneous slabs for Fermi-Eyges discussion. Pb/Lexan/air simulates the upstream end of a double scattering beam line.\label{fig:stack}}
\end{figure}

\subsection{The Basic Theory}
Consider Figure\;\ref{fig:stack}. Our eventual goal is to find the phase space distribution and related quantities at any depth $z$ if an arbitrary Gaussian beam enters at $z=0$. Let $z$ be the beam direction and $x,y$ transverse directions ($y$ is up and $x,y,z$ form a right handed frame). For the present, suppose a {\em single proton} enters the {\em first slab}, whose upstream face is at $z=0$, along the $z$ axis. In a very short but mathematically sophisticated paper, Eyges \cite{eyges} showed that the probability of finding the proton later at some $z>0$ with $x$ in $dx$ and $\theta$ in $d\theta$ is
\begin{equation}\label{eqn:Pxt}
P(x,\theta)\,dx\,d\theta\;=\;\frac{1}{2\pi\sqrt{B}}\;e^{\displaystyle{
  -\,\frac{1}{2}\frac{A_0x^2-2A_1x\theta+A_2\theta^2}{B}}}\,dx\,d\theta
\end{equation}
The $A_n$ are moments of $T$ namely
\begin{eqnarray}
A_0(z)&\equiv&\int_0^z\,T(u)\,du\label{eqn:A0}\\
A_1(z)&\equiv&\int_0^z(z-u)\,T(u)\,du\label{eqn:A1}\\
A_2(z)&\equiv&\int_0^z(z-u)^2\,T(u)\,du\label{eqn:A2}\\
\noalign{\noindent and}
B(z)&\equiv&A_0A_2\;-\;A_1^2\label{eqn:B}
\end{eqnarray}
From now on, we suppress the $z$ dependence, it being understood that the $A_n$ and any related quantities are functions of depth. 

$B$ is greater than zero if there is any scattering at all, as the following proof by Jette \cite{jette} shows. The first line is true by inspection for any $z>0$ because $T>0$ (the mean squared angle can only increase with $z$).
\begin{eqnarray}
0\;&<&A_0\int_0^zT(u)\Bigl((z-u)-\frac{A_1}{A_0}\Bigr)^2\;du\nonumber\\
&&=A_0\int_0^z\Bigl(T(z-u)^2-2T(z-u)\frac{A_1}{A_0}+T\bigl(\frac{A_1}{A_0}\bigr)^2\Bigr)\;du\nonumber\\
&&=A_0\Bigl(A_2-2A_1\frac{A_1}{A_0}+A_0\bigl(\frac{A_1}{A_0}\bigr)^2\Bigr)\;=\;A_2A_0-A_1^2\;=\;B\nonumber
\end{eqnarray}

To find the distribution of $\theta$ irrespective of $x$ we integrate $P(x,\theta)$ over $x$. Completing the square in Eq.\,(\ref{eqn:Pxt}) by using Eq.\,(\ref{eqn:complete}) with $u\doteq x$  we find
\begin{equation}\label{eqn:Pxtx}
P(x,\theta)\;=\;\frac{1}{2\pi\sqrt{B}}\;\;e^{\displaystyle{-\frac{1}{2}\frac{\theta^2}{A_0}}}\;\;
  e^{\displaystyle{-\frac{1}{2}\frac{A_0}{B}(x-\frac{A_1}{A_0}\theta)^2}}
\end{equation}
\[
P(\theta)\;=\;\int_{-\infty}^{\infty}P(x,\theta)dx\;=\;\frac{1}{2\pi\sqrt{B}}\;\;
  e^{\displaystyle{-\frac{1}{2}\frac{\theta^2}{A_0}}}\;\;
  \int_{-\infty}^{\infty}e^{\displaystyle{-\frac{1}{2}\frac{A_0}{B}(x-\frac{A_1}{A_0}\theta)^2}}dx
\]
The term $A_1\theta/A_0$, independent of $x$, is equivalent to a shift in origin which we can ignore since the integral runs from $-\infty$ to $\infty$. Using Eq.\,(\ref{eqn:PGnorm}) we find
\begin{equation}\label{eqn:Pt}
P(\theta)\;=\;\frac{1}{\sqrt{2\pi A_0}}\;e^{\displaystyle{-\frac{1}{2}\frac{\theta^2}{A_0}}}
\end{equation}
Comparing with Eq.\,(\ref{eqn:PG})we find that $A_0$ equals the variance of $\theta$
\begin{equation}\label{eqn:A0eqtsq}
A_0\;=\;<\theta^2>
\end{equation}
which was already obvious from Eq.\,\ref{eqn:A0} and the definition of $T$.

By virtue of the symmetry of Eq.\,(\ref{eqn:Pxt}) any equation in $\theta$ can be turned into a corresponding equation in $x$ by swapping $x,\theta$ and $A_0,A_2$. Thus Eq.\,(\ref{eqn:Pxtx}) becomes
\begin{equation}\label{eqn:Pxtt}
P(x,\theta)\;=\;\frac{1}{2\pi\sqrt{B}}\;\;e^{\displaystyle{-\frac{1}{2}\frac{x^2}{A_2}}}\;\;
  e^{\displaystyle{-\frac{1}{2}\frac{A_2}{B}(\theta-\frac{A_1}{A_2}x)^2}}
\end{equation}
while Eq.\,(\ref{eqn:Pt}) becomes
\begin{equation}\label{eqn:Px}
P(x)\;=\;\frac{1}{\sqrt{2\pi A_2}}\;e^{\displaystyle{-\frac{1}{2}\frac{x^2}{A_2}}}
\end{equation}
and $A_2$ equals the variance of $x$.
\begin{equation}\label{eqn:A2eqxsq}
A_2\;=\;<x^2>
\end{equation}
That is less obvious, but a simple physical argument will be given later (Section\;\ref{sec:PK}).

In addition to the gross variance of $x$ and $\theta$ we require $\theta_C^2$, the variance of $\theta$ at a given $x$ (the angular spread of protons emerging from a narrow slit at some $x$) and $x_\mathrm{eff}^2$, the variance of $x$ at a given $\theta$ (harder to visualize).\footnote{~The name $x_\mathrm{eff}$ is explained later.} According to probability theory the conditional probability of $\theta$ given $x$ is
\begin{eqnarray}
P(\theta|\,x)\,d\theta&=&\frac{P(x,\theta)\,dx\,d\theta}{P(x)\,dx}\\
&=&\sqrt{\frac{A_2}{2\pi B}}\;\;e^{\displaystyle{-\frac{1}{2}\frac{A_2}{B}(\theta-\frac{A_1}{A_2}x)^2}}\;d\theta\label{eqn:Pt|x}
\end{eqnarray}
showing that, given $x$, the most probable $\theta$ is
\begin{equation}\label{eqn:mptgx}
\theta\;=\;\frac{A_1}{A_2}\,x
\end{equation}
Furthermore
\begin{equation}\label{eqn:tcsq}
\theta_C^2\;=\;<\theta|_x^2>\;=\;\int_{-\infty}^{\infty}\theta^2\,P(\theta|\,x)\,d\theta\;=\;\frac{B}{A_2}
\end{equation}
$\theta_C$ is independent of $x$: the angular spread of protons emerging from a slit is independent of the tranverse position of the slit.\,\footnote{~That is somewhat obvious from first principles. Although the {\em mean} angle of protons emerging from a slit depends on its transverse position, the `scattering history' of those protons, in small angle approximation, is statistically the same as on-axis protons and therefore the {\em spread} about the mean is the same.} 

Similarly
\begin{equation}
P(x|\,\theta)\,dx\;=\;
  \sqrt{\frac{A_0}{2\pi B}}\;\;e^{\displaystyle{-\frac{1}{2}\frac{A_0}{B}(x-\frac{A_1}{A_0}\theta)^2}}\;dx\label{eqn:Px|t}
\end{equation}
showing that, given $\theta$, the most probable $x$ is
\begin{equation}\label{eqn:mpxgt}
x\;=\;\frac{A_1}{A_0}\,\theta
\end{equation}
and
\begin{equation}\label{eqn:xeffsq}
x_\mathrm{eff}^2\;=\;<x|_\theta^2>\;=\;\int_{-\infty}^{\infty}x^2\,P(x|\,\theta)\,dx\;=\;\frac{B}{A_0}
\end{equation}
which is independent of $\theta$. The transverse spread of protons at a given inclination to the axis is independent of the inclination. 

We have given simple physical interpretations of $A_0$ and $A_2$. To interpret $A_1$ consider $<x\,\theta>$, the covariance of $x$ and $\theta$. The average can be arranged in either of two ways. Choosing the outer average to be over $x$ (denoted $<>_x$) and denoting `$\theta$ given $x$' by $\theta|_x$ we have
\begin{eqnarray*}
<x\,\theta>_x&=&<x\times\,\theta|_x>_x\\
&=&<x<\theta|_x>_\theta>_x\\
&=&<x\left((A_1/A_2)x\right)>_x\\
&=&(A_1/A_2)<x^2>_x\;=\;(A_1/A_2)\,A_2\;=\;A_1
\end{eqnarray*}
The third line follows from Eq.\,(\ref{eqn:mptgx}) and the fact that the distribution of $\theta|_x$ is Gaussian so the mean equals the most probable value. The fourth line follows from Eq.\,(\ref{eqn:A2eqxsq}). Thus
\begin{equation}\label{eqn:A1eqxt}
A_1\;=\;<x\,\theta>
\end{equation}
Of course, the same result is obtained doing the averages the other way.

So far we have linked $A_0$ to the gross angular spread (Eq.\,\ref{eqn:A0eqtsq}), $A_1$ to the correlation between $x$ and $\theta$  (Eq.\,\ref{eqn:A1eqxt}) and $A_2$ to the gross transverse spread (Eq.\,\ref{eqn:A2eqxsq}). Further interpretations of the $A_n$ and $B$, of a more geometric sort, will appear in the next section. As an application of the ones given so far, consider
\begin{equation}
\theta_C^2\;=\;\frac{B}{A_2}\;=\;\frac{A_0A_2-A_1^2}{A_2}\;=\;A_0-\frac{A_1^2}{A_2}\;=\;\;<\theta^2>-\;\frac{<x\,\theta>^2}{<x^2>}
\end{equation}
The angular spread at a point (spread of protons emerging from a narrow slit) is less than the gross angular spread unless $<x\,\theta>\,=0$.

Note: many writers absorb the physical interpretation of the $A_n$ into the notation, writing Eq.\,(\ref{eqn:Pxt}) for instance as
\begin{equation}\nonumber
P(x,\theta)\,dx\,d\theta\;=\;\frac{1}{2\pi\sqrt{\overline{x^2}\;\overline{\theta^2}-\overline{x\theta}^2}}\;e^{\displaystyle{
  -\,\frac{1}{2}\frac{\overline{\theta^2}x^2-2\overline{x\theta}x\theta+\overline{x^2}\theta^2}{\overline{x^2}\;\overline{\theta^2}-\overline{x\theta}^2}}}\,dx\,d\theta
\end{equation}
We prefer Eyges' notation as being easier to read. It also emphasizes that the $A_n$ are simply three numbers that can be calculated as a function of $z$, the depth in the stack.

\begin{figure}[p]
\centering\includegraphics[width=3.75in,height=3.0in]{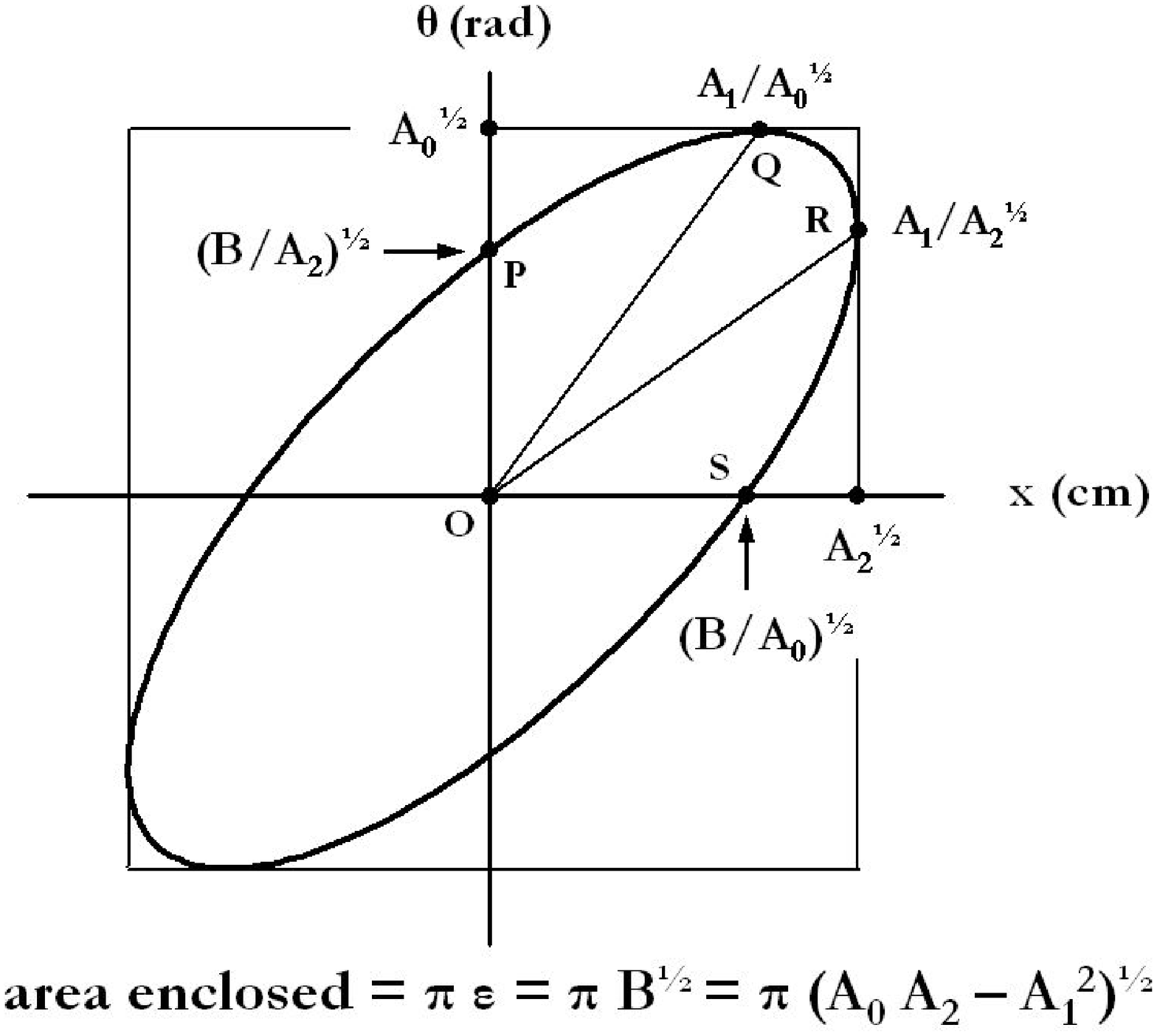} 
\caption{The beam ellipse in $A_n$, $B$ notation.\label{fig:ellipseA}}
\end{figure}
\begin{figure}[p]
\centering\includegraphics[width=3.75in,height=3.0in]{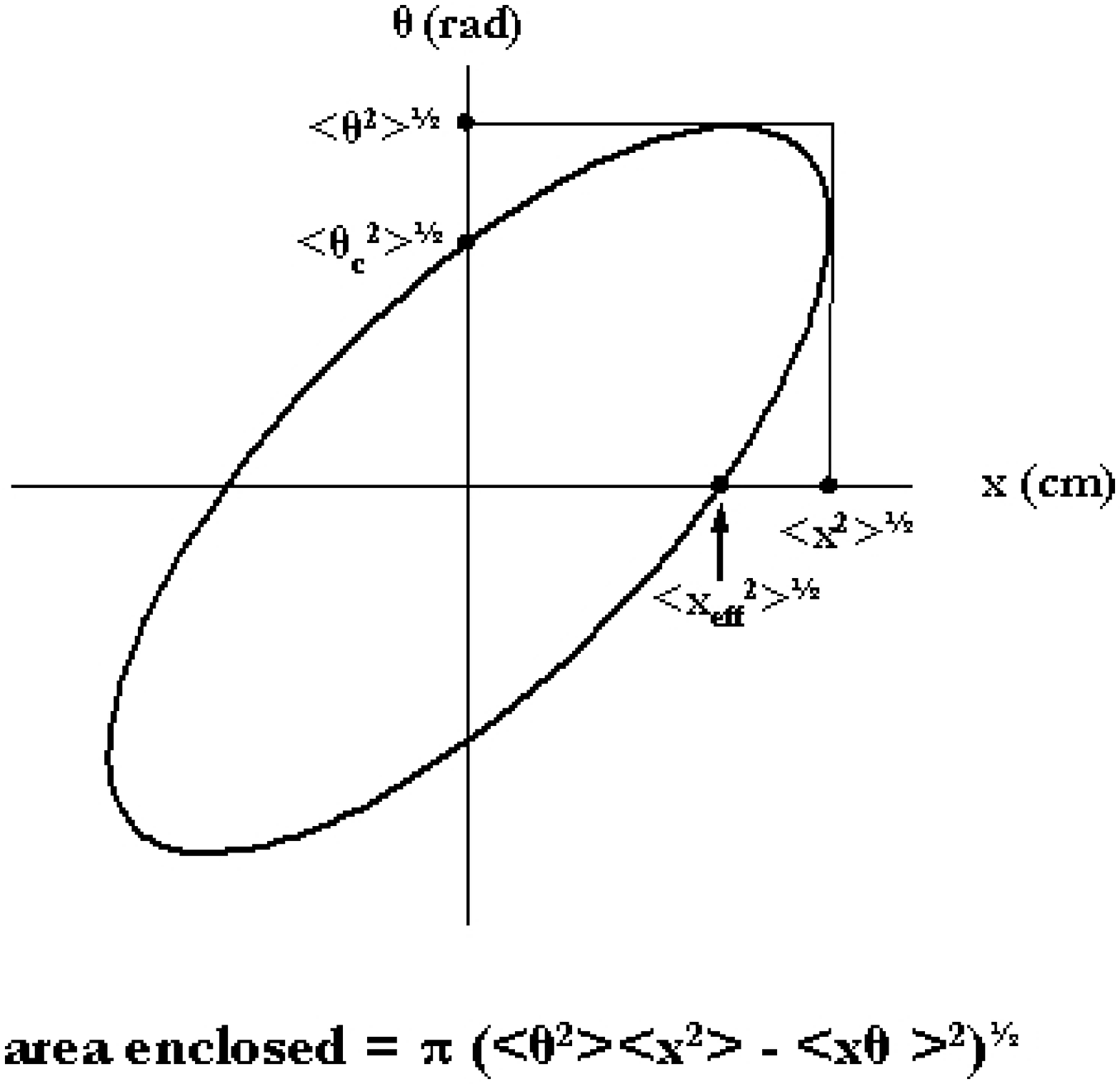} 
\caption{The beam ellipse in $x,\theta$ notation.\label{fig:ellipseT}}
\end{figure}

\subsection{The Beam Ellipse}\label{sec:beamEllipse}
The contour in phase space on which the probability density Eq.\,(\ref{eqn:Pxt}) equals $e^{-1/2}=0.606=61\%$ of its central value is found by setting
\begin{equation}\label{eqn:ellipseAB}
F(x,\theta)\;\equiv\;A_0x^2-2A_1x\theta+A_2\theta^2\;\doteq\;B
\end{equation}
According to the analytic geometry of conic sections Eq.\,(\ref{eqn:ellipseAB}) describes an ellipse because $B>0$. It is a centered ellipse because $F$ is invariant under $x\rightarrow-x,\;\theta\rightarrow-\theta$ (no linear terms in $x$ or $\theta$). 

The extrema $\hat{\theta}$ of the ellipse are found as usual by setting $dF/dx$ equal to 0, giving $x=A_1\theta/A_0$. Putting that into Eq.\,(\ref{eqn:ellipseAB}) and simplifying we find that
\begin{equation}\label{eqn:hatt}
\hat{\theta}\;=\;\pm\sqrt{A_0}\;=\;<\theta^2>^{1/2}\hbox{\quad occurring at\quad}x\;=\;\pm\,A_1/\sqrt{A_0}
\end{equation}
Similarly  
\begin{equation}\label{eqn:hatx}
\hat{x}\;=\;\pm\sqrt{A_2}\;=\;<x^2>^{1/2}\hbox{\quad occurring at\quad}\theta\;=\;\pm\,A_1/\sqrt{A_2}
\end{equation}
In other words, the rms values of $x$ and $\theta$ define the bounding box of the ellipse. Figure\,\ref{fig:ellipseA} shows what we have learned so far in $A,B$ notation. Figure\,\ref{fig:ellipseT} shows the same thing in $x,\theta$ notation. 

In Appendix \ref{sec:ellipse} we show that the area bounded by a centered ellipse in $x,y$ is
\[\pi\epsilon\;=\;\pi\;y_\mathrm{int}\;x_\mathrm{max}\;=\;\pi\;x_\mathrm{int}\;y_\mathrm{max}\]
According to Eq.\,(\ref{eqn:tcsq}) $\theta_C$ is the rms spread in $\theta$ at any given value of $x$, including 0, which means $\theta_C$ is the $y$ intercept of the ellipse. Eq.\,(\ref{eqn:A2}) gives the conjugate maximum of the ellipse, $x_\mathrm{max}=\sqrt{A_2}$. Thus
\begin{equation}\label{eqn:epsilon}
\epsilon\;=\;\theta_C<x^2>^{1/2}\;=\;\sqrt{B/A_2}\sqrt{A_2}\;=\;\sqrt{B}
\end{equation}
or
\begin{equation}
\hbox{area bounded by the ellipse}\;=\;\pi\;\sqrt{B}
\end{equation}
which is the interpretation of $B$. In accelerator physics the bounded area is called the {\em emittance} of the beam \cite{carey}.\footnote{~Confusingly, in the early accelerator literature area/$\pi$ (our $\sqrt{B}$) is sometimes called the emittance. If we quote the emittance as e.g. `$10\,\pi$\;mm\,mrad' it is clear we mean area, with $\sqrt{B}=10$\;mm\,mrad.} However, in accelerator physics the beam ellipse is merely a convenience (we know how to transport it through magnets and drifts) and Gaussian phase space density is usually only a rough approximation. By contrast, in proton transport through matter the phase space density really is Gaussian to a very good approximation and iso-density contours really are elliptical.
 
$A_1=\;<x\,\theta>$ is related to the tilt of the beam ellipse. It is conventional \cite{icru35} to derive an equation for $\Psi$, the angle from the $x$ axis to the principal axis of the ellipse, namely
\begin{equation}\label{eqn:tan2Psi}
\tan2\,\Psi\;=\;\frac{2\,A_1}{A_2-A_0}
\end{equation}
This must be used with discretion as the denominator warns us: $A_2$ and $A_0$ have different dimensions and cannot be added. $\Psi$ refers to the angle as measured in a particular drawing of the beam ellipse, and the $A_n$ in Eq.\,(\ref{eqn:tan2Psi}) are appropriately scaled versions of the real $A_n$.

From Eq.\,(\ref{eqn:Pt|x}) the most probable $\theta$ given $x$ is $\theta=(A_1/A_2)\,x$, a line which we can identify with {\bf OR} in Figure\,\ref{fig:ellipseA}. Similarly from Eq.\,(\ref{eqn:Px|t}) we can identify {\bf OQ} as the locus of the most probable $x$ given $\theta$.

\subsection{Drawing the Ellipse Given the Moments}
Instead of deriving $\Psi$ let us solve a related problem: Draw the ellipse corresponding to given $A_0$, $A_1$ and $A_2$. In a computer program, any ellipse is most conveniently drawn using the parametric form of Appendix \ref{sec:ellipse} whose equations Eq.\,(\ref{eqn:x}) and Eq.\,(\ref{eqn:y}) we can rewrite  
\begin{eqnarray}
x\;&=&\;\hat{x}\;\cos t\label{eqn:Aprime}\\
\theta\;&=&\;\hat{\theta}\;\sin(t-\delta)\label{eqn:Bprime}
\end{eqnarray}
$\hat{x}$ and $\hat{\theta}$ define the bounding box in graph units. To generate points on the ellipse, let $t$ range in steps from $0$ to $2\pi$. To find $\delta$ note that Eq.\,(\ref{eqn:ab}), in our present notation, yields
\[\cos\delta\;=\;\frac{\epsilon}{\sqrt{A_0}\sqrt{A_2}}\;=\;\frac{\sqrt{B}}{\sqrt{A_0}\sqrt{A_2}}\;=\;\frac{\sqrt{A_0A_2-A_1^2}}{\sqrt{A_0A_2}}\]
or
\begin{equation}\label{eqn:delta}
\delta\;=\;\sin^{-1}(A_1/\sqrt{A_0A_2})
\end{equation}
Unlike using $B$ (always positive) as the third ellipse parameter, this procedure preserves the sign information (converging or diverging beam) in $A_1$ because the principal range of $\sin^{-1}$ is $-\pi/2$ to $\pi/2$. To plot the ellipse in (mm, mrad) if $\sqrt{A_2},\,\sqrt{A_0}$ are in (cm, rad) use
\begin{eqnarray}
\hat{x}&=&10\;\sqrt{A_2}\nonumber\\
\hat{\theta}&=&1000\;\sqrt{A_0}\nonumber
\end{eqnarray}
in Eqs.\,(\ref{eqn:Aprime}) and (\ref{eqn:Bprime}).

It may be (cf. Figure\,\ref{fig:ModelPS}) that we do not know the $A_n$ explicitly but wish to draw an ellipse around a set of points obtained by Monte Carlo, or otherwise. In that case we find the statistical rms values of $x_i$ and $\theta_i$ and fit a line to $\theta_i(x_i)$ to find $<d\theta/dx>$. Then
\begin{eqnarray*}
\alpha&=&\tan^{-1}(<\theta^2>^{1/2}/<x^2>^{1/2})\\
\psi&=&\tan^{-1}(<d\theta/dx>)\\
\delta&=&\sin^{-1}(\tan(2\psi)/\tan(2\alpha))
\end{eqnarray*}
In Figure\,\ref{fig:ModelPS} we multiplied $<x^2>^{1/2}$ and $<\theta^2>^{1/2}$ by 2.5 to obtain the larger ellipse.

\begin{figure}[p]
\centering\includegraphics[width=4.72in,height=3.5in]{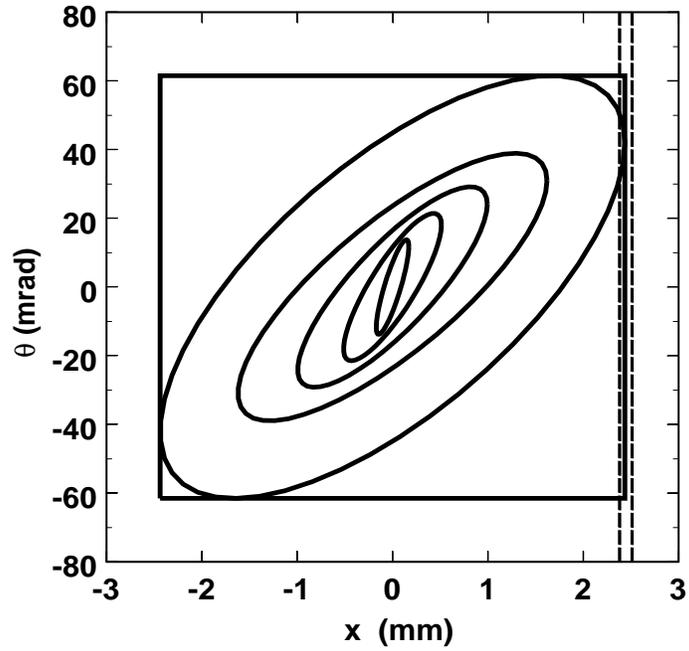} 
\caption{Evolution of the beam ellipse of an ideal 127\,MeV proton beam in near stopping thickness (0.97$\times$range) water divided into five equal slabs. The dashed lines represent the uncertainty range of a measurement by Preston and Koehler \cite{preston}.\label{fig:reality}}
\end{figure}
\begin{figure}[p]
\centering\includegraphics[width=4.72in,height=3.5in]{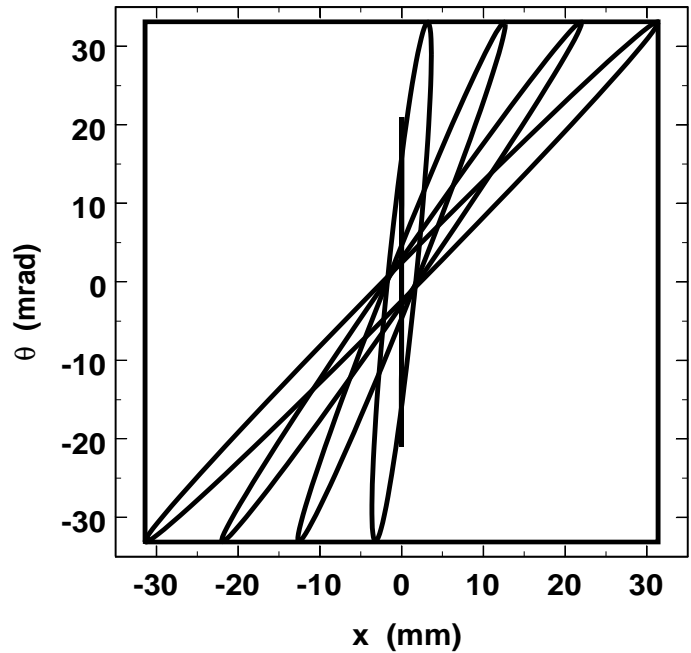} 
\caption{Evolution of the beam ellipse for a 230\,MeV proton pencil beam entering a Pb/Lexan/air (respectively 0.234/14.6/85.1\,cm) stack. The air is divided into 3 equal slabs.\label{fig:stack230ellipses}}
\end{figure}

\subsection{Ellipse Examples}
Figure\,\ref{fig:reality} is a reality check which puts to work everything we have learned so far. It shows beam ellipses corresponding to a 127\,MeV proton pencil beam propagating in a near stopping thickness (0.97$\times$range) slab of water, divided into five equal parts. We used $T_\mathrm{dM}$ for the scattering power. The bounding box of the final ellipse is shown. Its $x_{rms}$ agrees well with the measurement by Preston and Koehler \cite{preston}.

Next, for a problem of practical importance that can be fully handled by Fermi-Eyges theory, let's return to Figure\;\ref{fig:stack} which represents one step of a compensated range modulator \cite{BGcourse} and the air between it and the second scatterer. Anticipating the next section, Figure\,\ref{fig:stack230ellipses} shows the evolution of an ideal 230\,Mev incident proton beam in Pb/Lexan/air. The Pb approximates a thin scatterer. Only the distribution of $\theta$ changes. The Lexan introduces more scattering as well as some drift making the emittance non-zero. The air (here divided into three slabs) is an almost pure drift: no increase in $\theta_{rms}$ or emittance. Now let's see how we compute those ellipses. 

\subsection{Transporting the Beam Ellipse Through a Slab}
We want to generalize from a perfect beam entering a single slab to a beam of finite emittance entering a stack of slabs (each homogeneous and transversely infinite). Some slabs may be air (little scattering) or void (no scattering). We call this `mixed slab' or `stack' geometry.

Consider a perfect beam entering a homogeneous slab extending from 0 to $z_2$ with an imaginary break at $z_1$. Take (for instance) the second moment at $z_2$ namely 
\[\int_0^{z_2}(z_2-z')^2\,T(z')\,dz'\;=\;\int_0^{z_1}(z_2-z')^2\,T(z')\,dz'\;+\;\int_{z_1}^{z_2}(z_2-z')^2\,T(z')\,dz'\]
Consider the first integral on the RHS. If we add and subtract $z_1$ we may write it
\begin{eqnarray}
&&\int_0^{z_1}((z_2-z_1)+(z_1-z'))^2\,T(z')\,dz'\nonumber\\
&&=\int_0^{z_1}((z_2-z_1)^2+2(z_2-z_1)(z_1-z')+(z_1-z')^2)\,T(z')\,dz'\nonumber\\
&&=(z_2-z_1)^2\,A_{01}\;+\;2(z_2-z_1)\,A_{11}\;+\;A_{21}\nonumber
\end{eqnarray}
where $A_{01}$ stands for the zeroth moment at $z_1$ and so on. Thus
\[\int_0^{z_2}(z_2-z')^2\,T(z')\,dz'\;=
  \;(z_2-z_1)^2\,A_{01}\;+\;2(z_2-z_1)\,A_{11}\;+\;A_{21}\;+\;\int_{z_1}^{z_2}(z_2-z')^2\,T(z')\,dz'\]
Since $z_1$ and $z_2$ are perfectly general we can set $z_1$ to $0$ and $z_2$ to $z$. Proceeding similarly for the other two moments and letting unprimed denote a moment at $0$ and primed a moment at $z$ we find
\begin{eqnarray}
A_0'&=&A_0\;+\;\int_0^zT(z')\,dz'\label{eqn:a0tr}\\
A_1'&=&A_1\;+\;A_0\,z\;+\;\int_0^z(z-z')\,T(z')\,dz'\label{eqn:a1tr}\\
A_2'&=&A_2\;+\;2\,A_1\,z\;+\;A_0\,z^2\;+\;\int_0^z(z-z')^2\,T(z')\,dz'\label{eqn:a2tr}
\end{eqnarray}
There is no physics in Eqs.\,(\ref{eqn:a0tr}\,-\,\ref{eqn:a2tr}). They express a mathematical property of moments based on the additivity of definite integrals. The initial moments need not come from scattering. They may represent the parameters of an incident beam. Any set is valid provide $A_0\ge0$, $A_2\ge0$ and
\[B=A_0A_2-A_1^2\ge0\] 
If $A_1=\;<x\,\theta>$ is negative the beam ellipse slopes down ($\theta$ decreases as $x$ increases) which (recalling our discussion of phase space diagrams) describes a converging beam. That cannot be produced with scatterers, but can be, and often is, produced with magnets. After a sufficiently long drift, any converging beam will diverge as the protons cross the beam axis. That too is built into Eqs.\,(\ref{eqn:a0tr}\,-\,\ref{eqn:a2tr}). 

$z'$ stands for position along the stack. Its role in expressions like $(z-z')^2$ is obvious. In $T(z')$ it is more subtle. It is a bookeeping tool meaning `where we are in the stack'. As shown by Eq.\,(\ref{eqn:TdM}), one possible choice for $T$, we will need a kinematic quantity $pv$ (which depends on kinetic energy, which depends on $z'$) and atomic properties of the current material (which also depend on $z'$). The beam transport program will have to include a procedure which accepts $z'$ and, knowing the incident energy and how the stack is constructed, returns the variables upon which $T$ directly depends.\footnote{~In our implementation that subroutine, in module SPWRsubs.FOR, is called WhatsHere($z'$,\ \ $\ldots)$\;.}

To summarize this section, we have derived three equations which transport an initial set of $A$\,s (Fermi-Eyges moments or ellipse parameters) through a finite homogeneous slab. The initial $A$\,s can represent either an incident beam or previous scattering. If they are zero the incident beam is ideal and we recover the original definitions of the moments. The equations can be repeated as often as necessary to transport the beam through a stack. The kinetic energy, or some equivalent longitudinal quantity, must also be transported, of course.

\subsection{Emittance Change in a Drift}
If a slab is void $T(z')=0$, the integrals in Eqs.\,(\ref{eqn:a0tr}) through (\ref{eqn:a2tr}) vanish, and the remaining terms transport the incident ellipse through a drift. By direct calculation
\[B'\;=\;A_0'A_2'\,-\,{A_1'}^2\;=\;A_0A_2\,-\,{A_1}^2\;=\;B\]
for any $z$, recovering the fact that emittance is conserved in a drift (despite the fact that $A_1$, $A_2$, the phase space distribution, and the beam cross section all change, except in trivial cases). 

\subsection{Emittance Change in a Scatterer}
Eqs.\,(\ref{eqn:a0tr}\,-\,\ref{eqn:a2tr}) have another important consequence. Suppose a given beam (given $A$\,s) falls on a thin scatterer so that we can assume $T$ is nearly constant and evaluate the integrals accordingly. Then
\begin{eqnarray*}
A_0'&=&A_0+\tilde{T}\,\Delta z\\
A_1'&=&A_1+A_0\,\Delta z+\tilde{T}\,(\Delta z)^2/2\\
A_2'&=&A_2+2A_1\,\Delta z+A_0\,(\Delta z)^2+\tilde{T}\,(\Delta z)^3/3
\end{eqnarray*}
where $\tilde{T}$ is the value at the midpoint of $\Delta z$. To lowest order in $\Delta z$
\[
B'\;=\;A_0'A_2'-A_1'^2\;\simeq\;A_0A_2-A_1^2\;+\; \tilde{T}\,\Delta z\,A_2 
\]
or
\begin{equation}\nonumber
\Delta B\;=\;\Delta(\epsilon^2)\;=\;A_2\;(T\,\Delta z)\;=\;\,<x^2>\times\;\Delta<\theta^2>
\end{equation}
If we take the square root, the quadratic difference on the RHS is just $\theta_0$, the rms strength of the thin scatterer \cite{mcsbg}.
Thus the quadratic change in emittance in a thin scatterer is
\begin{equation}\label{eqn:epsThin}
\pi\sqrt{\epsilon_2^2-\epsilon_1^2}\;=\;\pi\;x_{rms}\;\theta_0
\end{equation}
where $x_{rms}$ is the rms size of the beam impinging on the scatterer. If a beam of known emittance enters $N$ scatterers separated by drifts, the final emittance is $\pi\epsilon$ where
\begin{equation}\label{eqn:epsBeam}
\epsilon^2\;=\;\epsilon_\mathrm{beam}^2\;+\;\sum_{i=1}^N\;x_{rms,\,i}^2\times\theta_{0,\,i}^2
\end{equation}

Thus, if a passive beam line can be approximated by thin scatterers separated by drifts (as is usually the case) the final emittance can be approximated without the full Fermi-Eyges treatment by just computing the beam size at each scatterer, which is easier (see Section\,\ref{sec:PK}). Of course, each $\theta_0$ must be computed with due regard for the proton energy at that scatterer.

Eq.\,(\ref{eqn:epsBeam}) has important consequences for beamline design. As Figures\,\ref{fig:DSbeam} and \ref{fig:DSphase} show, the transverse penumbra of a double scattered beam depends on the angular confusion at the patient collimator, $\theta_C$, which is proportional to $\epsilon$ (Eq.\,\ref{eqn:epsilon}). Therefore the penumbra can be reduced by placing the second scatterer further upstream, where the beam is smaller. (That is also obvious geometrically from Figure\,\ref{fig:DSbeam}.)

If the incident beam is small, the first scatterer (no matter how strong) will cause relatively little emittance increase. That accounts for the small penumbra of single scattered beams. In fact, their penumbra is usually dominated by the air (which acts as a second scatterer) between the first scatterer and the patient.

\subsection{Differential Form of the Transport Equations}
Broadly speaking there are two applications of transport theory: beamline design and dose reconstruction in the patient. Passive beamlines are composed of slabs (though some may be nonuniform) so Eqs.\,(\ref{eqn:a0tr}\,-\,\ref{eqn:a2tr}) are a good starting point. In the patient, however, material properties change voxel by voxel and a finer-grained approach is needed. We wish to compute the changes in moments as the beam passes through a voxel of length $\Delta z$, assuming the material is known. These are easily derived from Eqs.\,(\ref{eqn:a0tr}\,-\,\ref{eqn:a2tr}) by assuming $T$ approximately equals its central value $\tilde{T}$ in each voxel (`midpoint rule'). If, following Kanematsu \cite{kanematsu08}, we rewrite the result to minimize the number of multiplications we find
\begin{eqnarray}
\Delta A_0&=&\tilde{T}\Delta z\label{eqn:a0k}\\
\Delta A_1&=&(A_0+(\tilde{T}/2)\Delta z)\Delta z\label{eqn:a1k}\\
\Delta A_2&=&(2A_1+(A_0+(\tilde{T}/3)\Delta z)\Delta z)\Delta z\label{eqn:a2k}
\end{eqnarray}
These equations are usually embedded in a general procedure which also transports other quantities (e.g. kinetic energy or $pv$ or residual range) through $\Delta z$.

\subsection{Equivalent Sources}
It is sometimes convenient to mentally replace the beamline upstream of some measuring plane $z_\mathrm{MP}$ by an equivalent source. For example, we might want to replace the Pb/Lexan/air stack which represents one modulator step of a double scattering system by an equivalent point source some distance $S$ upstream of the location of the second scatterer. Knowing $S$ gives us the effective `throw' of the first scatterer, allowing us to adjust its strength to produce the desired Gaussian width on the second scatterer.

There are three different ways of defining an equivalent source. The first two are in ICRU\;Report\,35 \cite{icru35}. Our formulas agree with that reference, of course, but our derivations are more geometric. 

\subsubsection{Effective Extended Source}
\begin{figure}[p]
\centering\includegraphics[width=4.5in,height=3.92in]{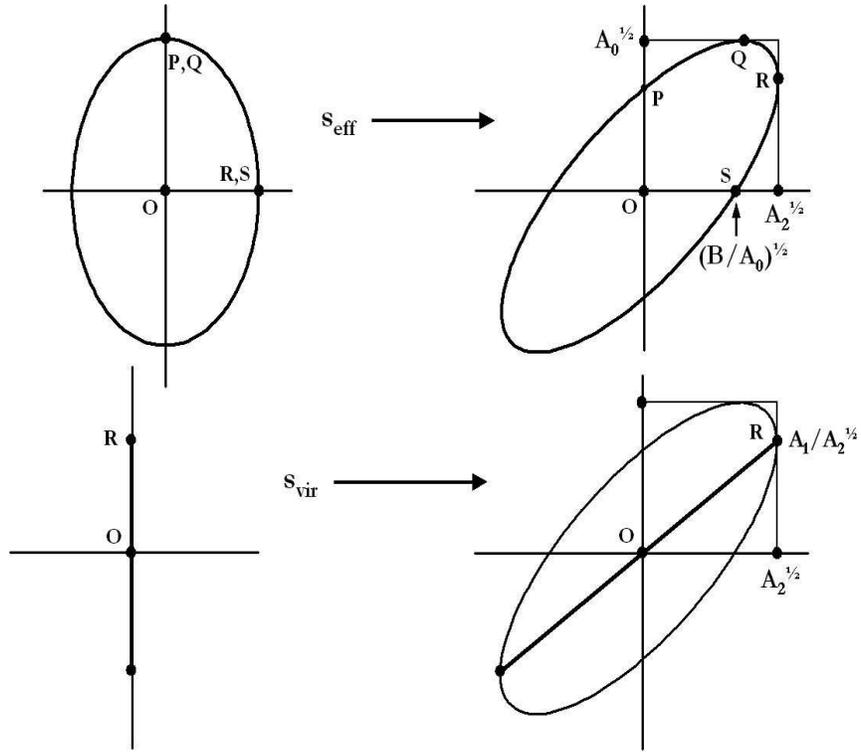} 
\caption{Ellipses for the (top) effective extended and (bottom) virtual point sources.\label{fig:sources}}
\end{figure}

The {\em effective extended source} is that ellipse which drifts into the given ellipse at $z_\mathrm{MP}$ in a drift distance $S_\mathrm{eff}$. Since there are infinitely many such ellipses, we require in addition that the source ellipse be erect, $<x\,\theta>\,=A_1=0$. Figure\,\ref{fig:sources} (top) illustrates the situation. Consider point {\bf Q}. The drift distance of a general phase space point $(x,\theta)$ from an initial $(0,\theta)$ is
\begin{equation}\label{eqn:S}
S\;=\;\frac{x}{\theta}
\end{equation}
From the known phase space coordinates of {\bf Q} (Figure\,\ref{fig:ellipseA}) we therefore find
\begin{equation}\label{eqn:Seff}
S_\mathrm{eff}\;=\;\frac{A_1/\sqrt{A_0}}{\sqrt{A_0}}\;=\;\frac{A_1}{A_0}\hbox{\qquad whence\qquad}z_\mathrm{eff}\;=\;z_\mathrm{MP}\;-\;\frac{A_1}{A_0}
\end{equation}
The effective extended source reproduces the given ellipse exactly. Emittance is the same since emittance does not change in a drift. Since points on the $x$ axis do not move in a drift, the rms size of the (erect) source ellipse is
\begin{equation}\label{eqn:xeff}
x_\mathrm{eff}\;=\;\sqrt{B/A_0}
\end{equation} 
as shown earlier (Eq.\,\ref{eqn:xeffsq}). That justifies the notation $x_\mathrm{eff}$.

\begin{figure}[p]
\centering\includegraphics[width=4.11in,height=3.5in]{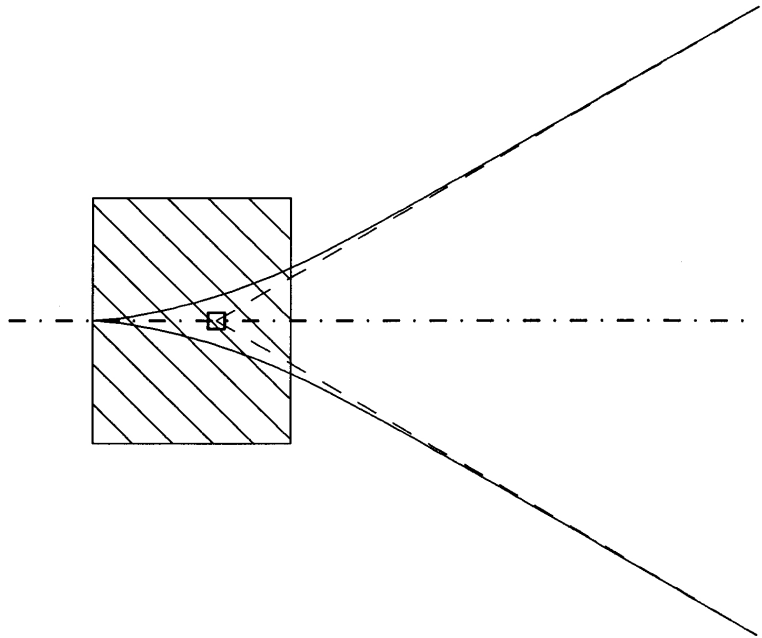} 
\caption{Asymptotic `effective scattering point' in a single slab.\label{fig:mcs4}}
\end{figure}

\begin{figure}[p]
\centering\includegraphics[width=4.54in,height=3.5in]{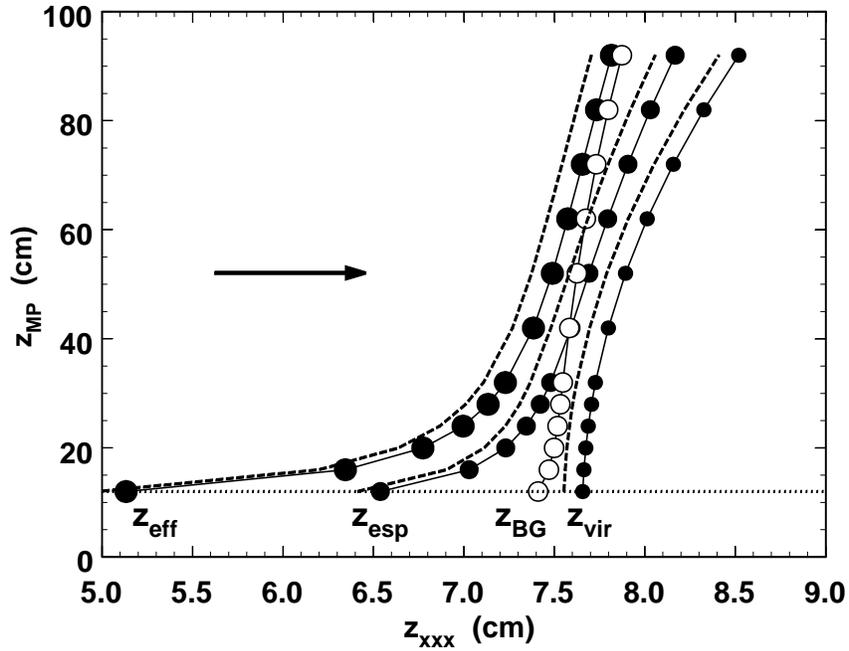} 
\caption{Source locations v. depth of measuring plane using a 12\,cm Lexan slab starting at $z=0$, followed by air, and an ideal 160\,MeV proton beam. Full circles use $T_\mathrm{dM}$\,; dashed lines use $T_\mathrm{IC}$ \cite{scatPower2010}\,; open circles ($z_\mathrm{BG}$) use Highland's formula \cite{mcsbg} (compare with $z_\mathrm{esp}$\,, full circles of same size).\label{fig:zeff160air}}
\end{figure}
\begin{figure}[p]
\centering\includegraphics[width=5.04in,height=3.5in]{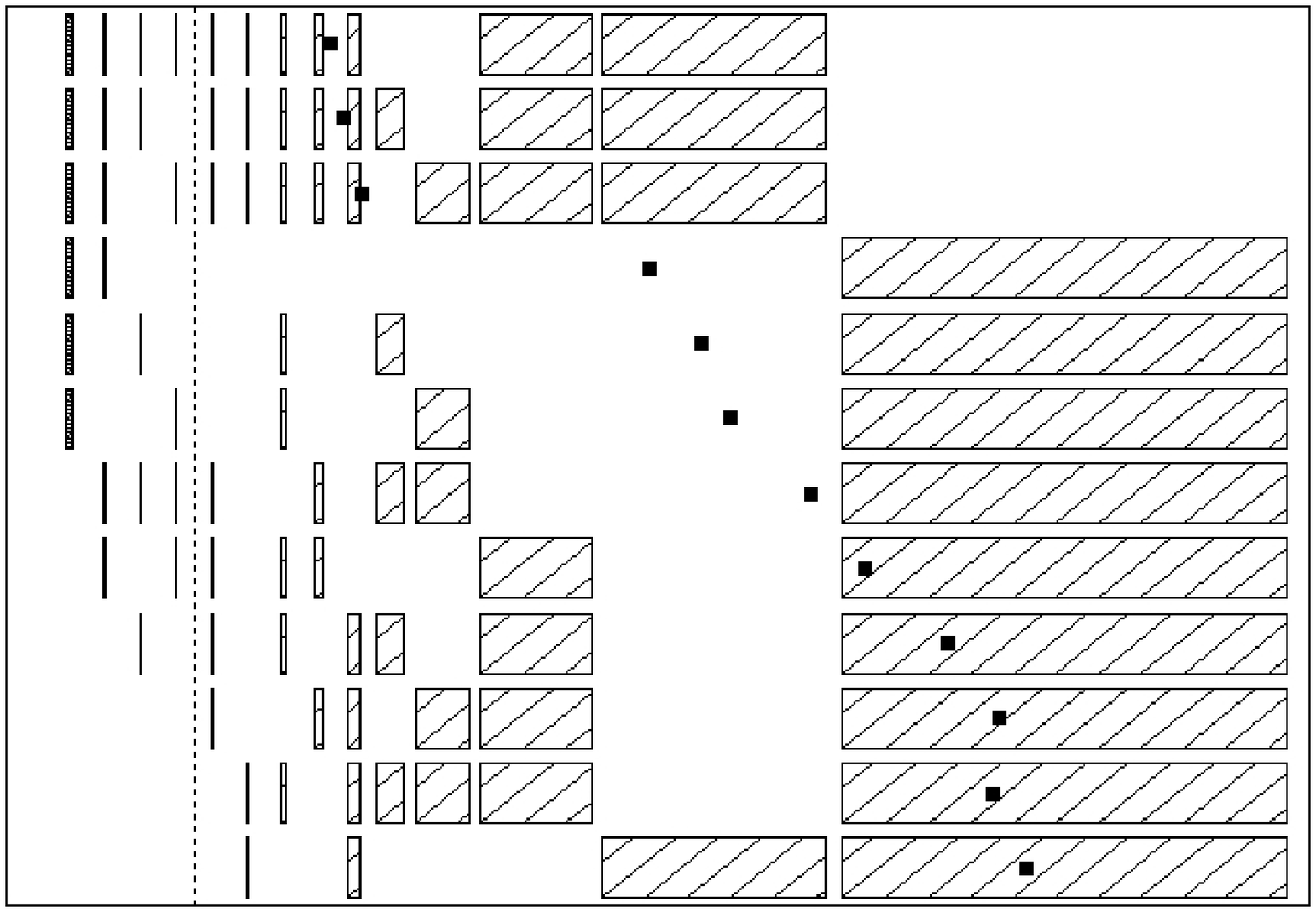} 
\caption{Practical application of the effective scattering point. The first scatterer of the Burr Center `STAR' beam uses Pb and Lexan degraders controlled by a computer to produce the desired SOBP. As the degraders change so does the effective scattering point (filled squares). Note the jump at the major binary transition. The related change in $1/r^2$ is compensated by adjusting the dwell time of each configuration in the beam.\label{fig:LOLz0}}
\end{figure}

\subsubsection{Virtual Point Source}
The {\em virtual point source} is the ellipse that would drift into the line OR, the locus of the most probable proton direction given $x$. Therefore it is the point from which the protons observed at $z_\mathrm{MP}$ seem to emanate. For a unique definition we again require that the source ellipse be erect (Figure\,\ref{fig:sources}, bottom). A line is a degenerate ellipse (emittance = $\sqrt{B}$ = 0) and an erect line is a point source with finite angular spread. Its parameters can be read from Figure\,\ref{fig:sources}. Using Eq.\,(\ref{eqn:S}) and the coordinates of {\bf R} we find
\begin{equation}\label{eqn:Svir}
S_\mathrm{vir}\;=\;\frac{\sqrt{A_2}}{A_1/\sqrt{A_2}}\;=\;\frac{A_2}{A_1}\hbox{\qquad whence\qquad}z_\mathrm{vir}\;=\;z_\mathrm{MP}\;-\;\frac{A_2}{A_1}
\end{equation}
The gross angular spread of the given ellipse is the quadratic sum of the spread of the virtual point source and the angular confusion:
\begin{equation}\overline{\theta^2}\;=\;\overline{\theta_{\mathrm{vir}}^2}\;+\;\theta_C^2\end{equation}
as can be verified by expressing each term in terms of the $A$'s.

\subsubsection{Effective Scattering Point}
The author, ignorant at that time of Fermi-Eyges theory, defined \cite{mcsbg} a third equivalent source distance using the back projected asymptote of $\sigma_x(z)$ (Figure\,\ref{fig:mcs4}). Kanematsu later called this the {\em effective scattering point} \cite{kanematsuLD}. It can be identified with the upper right corner of the bounding box. Using Eq.\,(\ref{eqn:S}) as before we obtain
\begin{equation}\label{eqn:Sesp}
S_\mathrm{esp}\;=\;\frac{\sqrt{A_2}}{\sqrt{A_0}}\;=\;\sqrt{\frac{A_2}{A_0}}\hbox{\qquad whence\qquad}z_\mathrm{esp}\;=\;z_\mathrm{MP}\;-\;\sqrt{\frac{A_2}{A_0}}
\end{equation}
Looking at Figure\,\ref{fig:ellipseA} and keeping Eq.\,(\ref{eqn:S}) in mind we can immediately write
\begin{equation}
S_\mathrm{vir}>S_\mathrm{esp}>S_\mathrm{eff}\hbox{\qquad whence\qquad}z_\mathrm{vir}<z_\mathrm{esp}<z_\mathrm{eff}
\end{equation} 
After any substantial drift $B$, though it remains constant, is much smaller than its largest possible value $A_0A_2$ (see Figure\,\ref{fig:stack230ellipses}). The three source distances become very nearly equal, and the potentially nettlesome question of which one to use in a given problem becomes moot.

\subsubsection{Examples}
Suppose we have a system not unlike the Burr Center eye treatment line, a 160\,MeV proton beam traversing 12\,cm of Lexan followed by $\approx90$\,cm of air. (The residual range is 3.9\,cm H$_2$O.) Figure\,\ref{fig:zeff160air} shows various computations of the equivalent source as a function of the depth of the measuring plane (MP). Full lines, full circles are computed with the preferred scattering power $T_\mathrm{dM}$; dashed lines are computed with $T_\mathrm{IC}$\,, the same as $T_\mathrm{dM}$ without the single scattering correction \cite{scatPower2010}. The difference is minor. Open circles represent the method described in \cite{mcsbg} (back projection of the beam envelope at the $1\sigma$ level) which we often use in place of Fermi-Eyges theory. In principle it should be compared with $z_\mathrm{esp}$, but it uses Highland's formula, making the comparison questionable.

Air has a noticeable effect on the calculation. The emittance increase in air is $\approx2.4\times$. If we substitute helium or hydrogen, the emittance increase is much less and the curves rise almost vertically (little change in $z_\mathrm{xxx}$ with $z_\mathrm{MP}$).

All that said, the real message of Figure\,\ref{fig:zeff160air} is how little difference any of this makes. The full range of source positions, once we are some distance downstream, is a few mm, and the total spread in the distance of the source to the MP, over all methods, is around $1\%$.

Figure\,\ref{fig:LOLz0} shows a practical application. The Burr Center neurosurgery beam is a single scattered beam using binary sets of Pb and Lexan slabs, manipulated by a computer, to scatter and range modulate the beam. As the slab configuration changes, so does the effective origin of protons, resulting in a small change in fluence. That change is compensated, to maintain a flat spread-out Bragg peak, by adjusting the in-beam time of each configuration. For Figure\,\ref{fig:LOLz0}, the equivalent source was computed according to the $z_\mathrm{BG}$ (effective scattering point) method \cite{mcsbg}. The others would give very similar source distances as we have just seen.  

\subsection{Beam Fraction Contained in a Generalized Beam Ellipse}  
The fraction of beam contained in the beam ellipse is of interest if, for instance, a degraded beam serves as the source for a magnetic beam line. It is obtained by integrating that part of the probability density Eq.\,(\ref{eqn:Pxt}) that lies within the ellipse. We can do that in closed form by transforming the ellipse into an erect ellipse (denoted with a prime) and then into a circle (double prime), while the number of phase space points inside remains the same. 

Let us first define the ellipse more generally than we did in Eq.\,\ref{eqn:ellipseAB}, letting
\begin{equation}\label{eqn:ellipseRho}
A_0x^2-2A_1x\theta+A_2\theta^2\;\doteq\;\rho^2\,B
\end{equation}
where $\rho>0$ is a dimensionless parameter. Following the methods of Sec.\,\ref{sec:beamEllipse}, this ellipse has bounding box $\hat{x}=\rho\sqrt{A_2}$\;, $\hat\theta=\rho\sqrt{A_0}$ and area $\pi\rho^2\sqrt{B}$. 

The effective extended source ellipse corresponding to the phase space probability density of Eq.\,\ref{eqn:Pxt} is erect by definition. It has $B'=B$ (emittance conserved in a drift) and $A'_0=A_0$ (no scattering in a drift) but  $A'_2=x_\mathrm{eff}^2=B'/A_0'=B/A_0$, so in terms of the original $A$s
\begin{equation}\nonumber
P'(x,\theta)\,dx\,d\theta\;=\;\frac{1}{2\pi\sqrt{B}}\;e^{\displaystyle{
  -\,\frac{1}{2}\frac{A_0x^2+B\theta^2/A_0}{B}}}\,dx\,d\theta
\end{equation}
and the generalized beam ellipse is now erect with bounding box $\hat{x}'=\rho\sqrt{B/A_0}$\;, $\hat{\theta}'=\hat{\theta}=\rho\sqrt{A_0}$ and area $\pi\rho^2\sqrt{B}$. No phase space points cross the ellipse during a drift, so the beam fraction contained in the new ellipse is the same as before. 

Now define variables $u,v$ such that the ellipse in $u,v$ space is a circle: $u=\sqrt{A_0}x,\;v=\sqrt{B/A_0}\;\theta$. Let $u^2+v^2=r^2$ and replace $du\,dv$ by $r\,dr\,d\phi$. Then
\begin{equation}
P''(r,\phi)\,r\,dr\,d\phi\;=\;\frac{1}{2\pi\sqrt{B}}\;e^{\displaystyle{
  -\,\frac{1}{2}\frac{r^2}{B}}}\,r\,dr\,d\phi
\end{equation}
The ellipse, now a circle of radius $\rho\sqrt{B}$, still contains the same beam fraction and $P''$ is still normalized cf. Eq.\,(\ref{eqn:PGcylnorm}). Therefore the desired beam fraction is
\begin{equation}\label{eqn:frho}
f_\rho\;=\;\int_0^{2\pi}\int_0^{\rho\sqrt{B}} P''(r,\phi)\,r\,dr\,d\phi\;=\;1- e^{\displaystyle{-\,\frac{1}{2}\rho^2}}
\;=\;1-\frac{P''(\rho\sqrt{B},\phi)}{P''(0,\phi)}
\end{equation}
The last equality says that if we take the ellipse as the contour where the probability density is down by e.g. $39.3\%$ from its central value (corresponding to $\rho=1$, the `standard' ellipse), then the fraction of beam inside is also $39.3\%$. We have simply generalized to phase space a property of cylindrical Gaussians pointed out by Preston and Koehler \cite{preston}.

Alternatively, one might wish to find the value of $\rho$ and the corresponding emittance of an ellipse containing a stipulated fraction $f_\rho$ of the beam. From Eq.\,(\ref{eqn:frho})
\begin{equation}\label{eqn:rho}
\rho^2\;=\;-2\ln(1-f_\rho)
\end{equation}
corresponding to emittance $\pi\rho^2\sqrt{B}$. 

For example, an ideal 230\,MeV proton beam traverses a 10.7\,cm carbon degrader (1.8\;g/cm$^3$), emerging with 150.2\,MeV and emittance $\pi\sqrt{B}=19.63\,\pi$\,mm\,mrad and entering a magnetic beam transport, whose designer defines the emittance ellipse as containing $95\%$ of the beam. What is the emittance from her point of view? From Eq.\,(\ref{eqn:rho}), $\rho^2=5.99$ so the 2D emittance $\pi\rho^2\sqrt{B}=118$\,mm\,mrad, six times the `Fermi-Eyges' value. This happens in both transverse directions, and the energy distribution is also smeared out, so the transmission of a degrader/re-analyzer system is very small when the energy step is large.

\subsection{Summary}
\begin{quote}
The joint probability of $x$ and $\theta$ (projected angle) at any depth $z$ in a homogeneous slab has a Gaussian form (Eq.\,\ref{eqn:Pxt}) involving three functions $A_n(z)$ which are moments of the scattering power $T$ (Eqs.\,\ref{eqn:A0}\,-\,\ref{eqn:B}).

The combination $B\equiv A_0A_2-A_1^2$ is positive if there is any scattering at all.

The distribution of $\theta$ irrespective of $x$ is Gaussian with variance $A_0$ (physical meaning of $A_0$, obvious from the definition of $T$).

The distribution of $x$ irrespective of $\theta$ is Gaussian with variance $A_2$ (physical meaning of $A_2$, will be more obvious later).

The rms spread of protons emerging from a slit at $z$ about their mean angle is the {\em angular confusion} $\theta_C=\sqrt{B/A_2}$. It is independent of $x$ (distance of the slit from the beam axis).

The quantity conjugate to $\theta_C$ is $x_\mathrm{eff}=\sqrt{B/A_0}$\,, called $x_\mathrm{eff}$ because it is the rms size of the effective extended source.

The physical meaning of $A_1$ is the covariance of $x$ and $\theta$ or $<x\,\theta>$.

$A_0$, $A_2$ and $A_1$ (or $B$) can be regarded as the three parameters of the beam ellipse at $z$. The bounding box of the ellipse is given by $\sqrt{A_2}$ and $\sqrt{A_0}$. The  tilt of the ellipse is related to $A_1$. Alternatively, the area enclosed by the ellipse  (beam emittance) is $\pi\sqrt{B}$ (physical meaning of $B$).

Therefore, given the three moments, the ellipse can be drawn (Eqs.\,\ref{eqn:Aprime}\,-\,\ref{eqn:delta}). 

Applying the theory developed so far to a 127\,MeV proton beam stopping in water, we find, at the maximum depth, an rms beam size $\sqrt{A_2}$ that agrees with experiment (Figure\,\ref{fig:reality}).

If a beam enters a finite homogeneous slab with moments $A_n$ its outgoing moments are $A'_n$\,, given by Eqs.\,(\ref{eqn:a0tr}\,-\,\ref{eqn:a2tr}). These equations can be iterated to transport an arbitrary (Gaussian) initial beam through mixed slabs. 

Eqs.\,(\ref{eqn:a0tr}\,-\,\ref{eqn:a2tr}) imply that, in a drift, emittance is conserved (Liouville Theorem), as is $A_0$. 

Eqs.\,(\ref{eqn:a0tr}\,-\,\ref{eqn:a2tr}) further imply that, in a thin scatterer, emittance increases by Eq.\,(\ref{eqn:epsThin}) (quadratic change $=\pi\times$\;rms beam size\;$\times$\;scatterer strength) and $A_2$ is conserved.

Eqs.\,(\ref{eqn:a0tr}\,-\,\ref{eqn:a2tr}) can be written in a differential form (Eqs.\,\ref{eqn:a0k}\,-\,\ref{eqn:a2k}) suitable for disordered heterogeneities (voxel-by-voxel treatment).

It is sometimes useful to replace the entire beam line under consideration by an equivalent source drifting to the depth of interest {\em in vacuo}. Three such sources are defined: 

The {\em effective extended source} is that erect ellipse which drifts into the ellipse at the POI. It exactly replaces the actual beam. 

The {\em virtual point source} is the point source from which the protons at the POI appear to emanate. 

The {\em effective scattering point} is the intercept of the back projected asymptote of $\sigma_x(z)$ with the beam axis. 

The first two can be identified with intersections of the beam ellipse with its bounding box, while the last is the upper right hand corner of the box. Various equations can be written for parameters and positions of the sources. After any substantial drift in air or {\em in vacuo} the source distances from the measuring plane converge (Figure\,\ref{fig:zeff160air}). 

The beam ellipse as defined conventionally  (exponent in Eq.\,\ref{eqn:Pxt} equals $-1/2$) contains $1-\exp(-1/2)\approx40\%$ of the beam. If a different definition of emittance is required, Eqs.\,(\ref{eqn:frho}) and (\ref{eqn:rho}) allow us to satisfy it. 

\end{quote}

\section{Beam Spreading in Matter}
As a simple but useful application of Fermi-Eyges theory we consider, in some detail, the transverse spreading of an ideal beam of protons or heavier ions in a homogeneous degrader. If the degrader happens to be water, that sets the limit on the smallness (as a function of depth) of targets that can be treated conformally in the ideal case. In practice, conformality will be worse because of non-ideal beams, heterogeneities, and non-elastic nuclear interactions.

\subsection{Theory of Preston and Koehler}\label{sec:PK}
During the first decade of proton radiotherapy at the Harvard Cyclotron Laboratory (HCL), William Preston and Andreas M. Koehler studied, theoretically and experimentally, the limits imposed by multiple Coulomb scattering (MCS) on proton beams of very small initial cross section (`pencil beams' in current terminology). Unfortunately their manu\-script \cite{preston} (hereinafter PK)  was turned down.\footnote{~Years later Andy told me the journal (he could not remember which one) had deemed proton radiotherapy of insufficient general interest. The paper was a decade or two ahead of its time.}

\begin{figure}[h]
\centering\includegraphics[width=4.41in,height=3.5in]{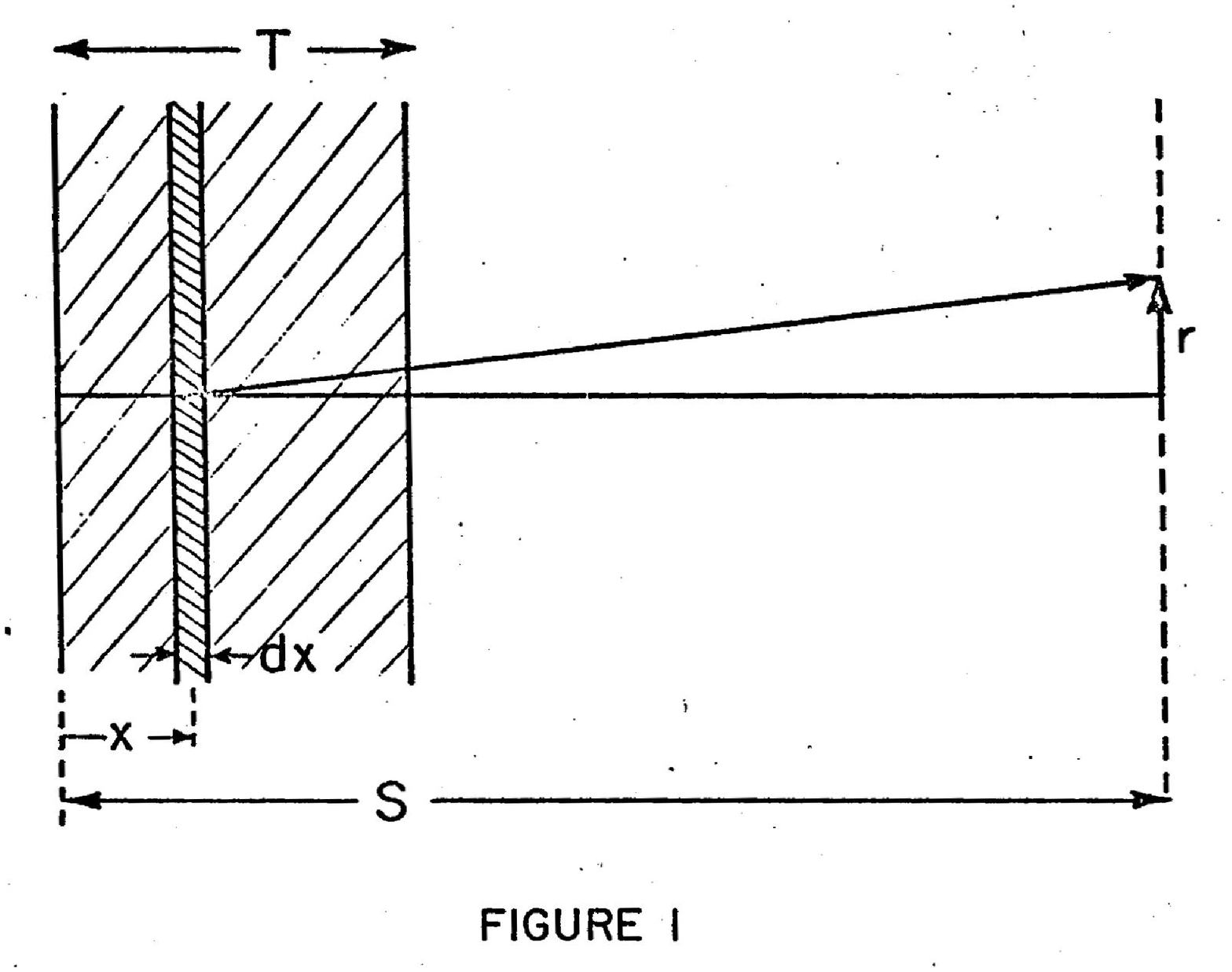} 
\caption{The construction of Preston and Koehler. The resultant displacement in the measuring plane equals the random sum of displacement vectors from infinitesimal elements of the slab.\label{fig:pk1}}
\end{figure}

In reading their paper one should realize that, though Fermi-Eyges theory already existed, they were unaware of it. Instead, they suggested an intuitive way of computing beam spreading,  essentially deriving Eq.\,(\ref{eqn:A2}) by direct physical reasoning. Using that model with the MCS theory of Bethe and Ashkin \cite{ashkin} they showed that the transverse spread at end-of-range of a proton beam stopping in any material is very nearly proportional to the range of the incident beam. The constant of proportionality depends on the material. They further showed that transverse beam spreading as a function of depth in any material obeys a universal law, for which they gave an analytic expression. That is, the basic shape of $\sigma_x$ of the pencil beam vs. depth, expressed in appropriate reduced variables, is independent of the stopping material, the incident energy and (as we will show later) the properties of the stopping ion.\footnote{~A third important result (which we will not cover here) was their demonstration that, as the initial beam cross section is made smaller, the Bragg peak vanishes. The diminution of fluence on axis due to MCS eventually outweighs the increase in proton stopping power. In current language, transverse equilibrium is lost.} Let us re-derive these results in the language of Fermi-Eyges theory, closely following their reasoning. 

Figure\,\ref{fig:pk1} is a facsimile of PK's construction which makes Eq.\,(\ref{eqn:A2}) (recalling $A_2=\,<x^2>$) intuitively obvious. Consider the random kick, in a scatterer element $dx$, projected onto a measuring plane (MP) at $S$ which can be anywhere downstream of the origin. A single kick has projected length $(S-x)\Delta\theta$. The rms spread in the MP is the quadratic sum of such kicks. Passing from the sum to an integral over $x$, introducing the definition of scattering power, and changing to our present notation ($x$ becomes $z$) we obtain Eq.\,(\ref{eqn:A2}). PK's results are all derived from this single equation.

PK used a scattering power $T$ derived from the MCS theory of Bethe and Ashkin \cite{ashkin}. We will instead use our `differential Moli\`ere' \cite{scatPower2010} scattering power (Eqs.\,\ref{eqn:TdM},\,\ref{eqn:fdM}).\footnote{~Any $T$ except $T_\mathrm{FR}$ yields almost identical results for water-like materials if only the transverse size of the beam at a given depth is considered \cite{scatPower2010}.} For a homogeneous slab, with $z$ any depth $\le R_1$, Eq.\,(\ref{eqn:A2}) then becomes 
\begin{equation}\label{eqn:x2z}
<x^2>_z\;=\;\frac{E_s^2\,z_\mathrm{I}^2}{X_S}\,\tilde{f}_\mathrm{dM}\int_0^z\frac{(z-z')^2}{(pv)^2}\,dz'
\end{equation}
(To avoid repetition we have anticipated heavy ions by writing a factor $z_\mathrm{I}$, the ion charge number, with every $E_s$.)

\begin{figure}[p]
\centering\includegraphics[width=4.50in,height=3.5in]{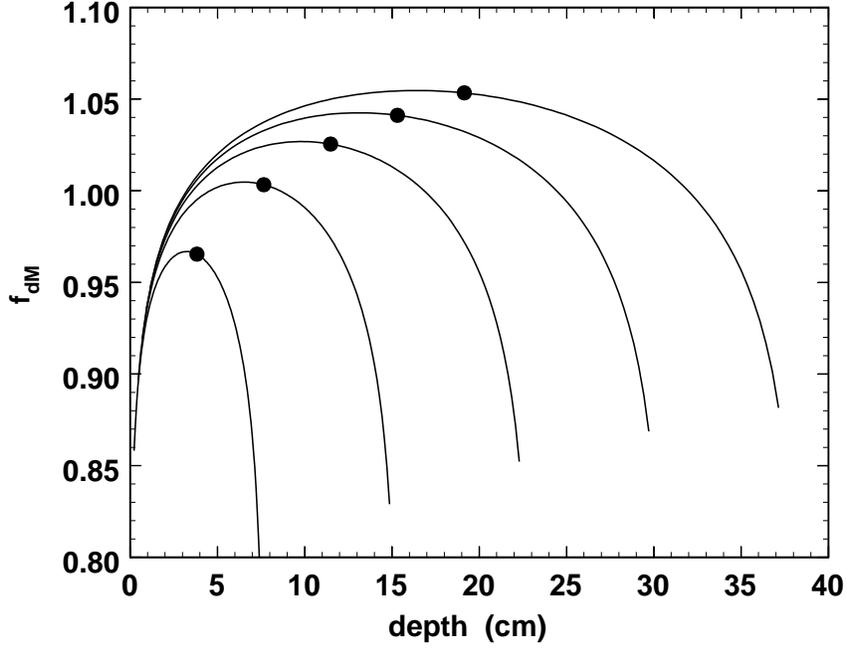} 
\caption{Scattering power correction factor $f_\mathrm{dM}$ for protons of five equally spaced ranges $R_1$ (the largest corresponding to 250\,MeV incident) stopping in water. The effective value $f_\mathrm{dM,\,z/2}$ used in evaluating Eq.\,\ref{eqn:x2R1Overas} at each range is marked by a full circle.\label{fig:fdMwater}}
\end{figure}
\begin{figure}[p]
\centering\includegraphics[width=4.55in,height=3.5in]{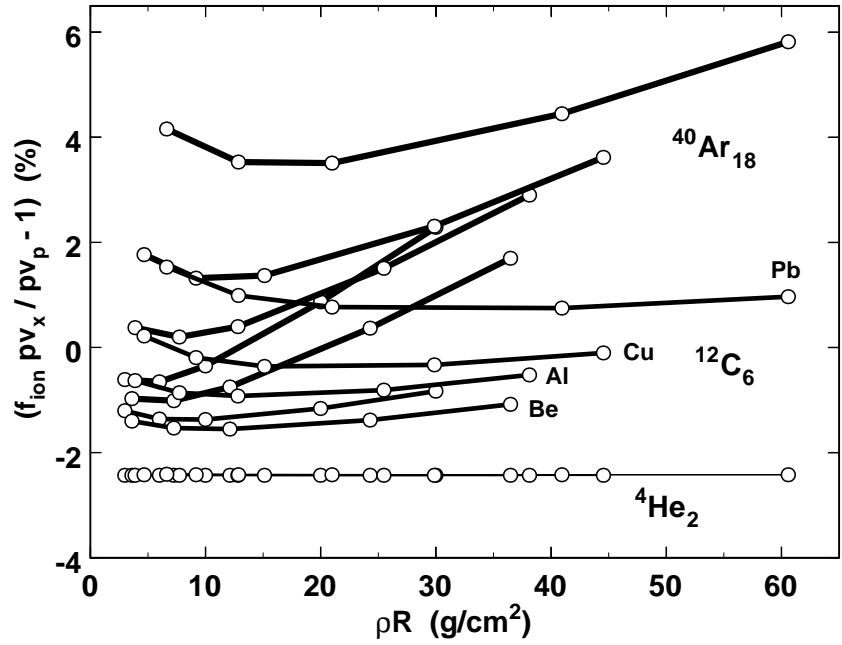} 
\caption{Accuracy of $pv$ reduction factor $f_\mathrm{I}=1./(z\sqrt{a})^{1.018}$ for three ions. Points are computed at 3, 6, 10, 20, 30\,cm water and equivalent ranges (same proton energy) in Be, Al, Cu and Pb.\label{fig:PVratio}}
\end{figure}

Though the scattering power is integrable numerically with  $f_\mathrm{dM}(p_1v_1,\,pv)$ included, to obtain results in closed form we have taken that slowly varying function outside the integral as an effective value $\tilde{f}_\mathrm{dM}$. In analogy to PK, let that be
\begin{equation}
\tilde{f}_\mathrm{dM}\;=\;f_\mathrm{dM}(p_1v_1,\,pv(z/2))
\end{equation}
that is to say, its value at half the depth of interest, call it $f_\mathrm{dM,\,z/2}$. Figure\,\ref{fig:fdMwater} shows the quality of that approximation. 

To integrate Eq.\,(\ref{eqn:x2z}) we express the kinematic quantity $pv$ in terms of $z'$ using the excellent approximation\footnote{~Nowadays this is known as the {\O}ver{\aa}s approximation \cite{overas}. Evidently PK were independently aware of it and the `weak {\O}ver{\aa}s' approximation used later.}
\begin{equation}\label{eqn:Overas}
(pv)^2\;=\;a(\rho R)^b\;=\;a\rho^b(R_1-z')^b\;=\;a\,\rho^{1+k}\,(R_1-z')^{1+k}
\end{equation}
where $a$ and $k$ depend on the stopping material. $k$ varies inversely with scattering length $X_S$ (or radiation length $X_0$) but is always small compared to 1 (Table\,\ref{tbl:Overas}). Eq.\,(\ref{eqn:x2z}) becomes
\begin{equation}\label{eqn:x2zOveras}
<x^2>_\mathrm{z}\;=\;\frac{E_s^2\,z_\mathrm{I}^2}{X_S}\,f_\mathrm{dM,\,z/2}\,\frac{1}{a\rho^{1+k}}\int_0^z\frac{(z-z')^2}{(R_1-z')^{1+k}}\;dz'
\end{equation}
If we desire the maximum (final) pencil beam size we set $z=R_1$ and, integrating, obtain
\begin{equation}\label{eqn:x2R1Overas}
<x^2>_\mathrm{R_1}\;=\;\frac{E_s^2\,z_\mathrm{I}^2}{X_S}\,f_\mathrm{dM,\,R_1/2}\;\frac{1}{a\rho^{1+k}}\,\frac{R_1^{\;2-k}}{2-k}
\end{equation}
A slightly less good approximation is obtained by setting $k\doteq0$ (`weak' {\O}ver{\aa}s approximation). In that case, $pv$ should be refit as:
\begin{equation}\label{eqn:weakOveras}
(pv)^2\;=\;A\,\rho\,(R_1-z')
\end{equation}
and Eq.\,(\ref{eqn:x2R1Overas}) becomes
\begin{equation}\label{eqn:x2R1weakOveras}
<x^2>_\mathrm{R_1}\;=\;\frac{E_s^2\,z_\mathrm{I}^2}{X_S}\,f_\mathrm{dM,\,R_1/2}\;\frac{1}{2A\,\rho}\;R_1^{\;2}
\end{equation}
This is PK's first result: $\sigma_x$ at end-of-range for a given material is nearly proportional to range. If we evaluate $A$ at $R_1/2$ and set $\sqrt{f_\mathrm{dM,\,R_1/2}}\approx1.015$ equal to 1 we are left with the very simple
\begin{equation}\label{eqn:sigxoR1simple}
\frac{\sigma_x(R_1)}{R_1}\;=\;\frac{E_s\,z_\mathrm{I}}{2\,(pv)_\mathrm{R_1/2}}\sqrt{\frac{R_1}{X_S}}
\end{equation}
which despite appearances is nearly independent of $R_1$. For protons in water with $R_1=5$\,cm it gives $2.37\%$\;, for $30$\,cm it gives 2.22\%. The more complicated procedure based on Eq.\,(\ref{eqn:x2R1Overas}) (Table\,\ref{tbl:Overas}) gives $2.25\%$. The three are equal for practical purposes.

Continuing PK's derivation, in the weak {\O}ver{\aa}s approximation Eq.\,(\ref{eqn:x2z}) becomes
\begin{equation}\label{x2zweakOveras}
<x^2>_\mathrm{z}\;=\;\frac{E_s^2\,z_\mathrm{I}^2}{X_S}\,f_\mathrm{dM,\,z/2}\,\frac{1}{A\rho}\int_0^z\frac{(z-z')^2}{R_1-z'}\;dz'
\end{equation}
which can be integrated giving
\begin{equation}\label{eqn:x2zweakOveras}
<x^2>_\mathrm{z}\;=\;\frac{E_s^2\,z_\mathrm{I}^2}{X_S}\,f_\mathrm{dM,\,z/2}\,\frac{R_1^{\;2}}{A\rho}\left[(1-t)^2\ln\left(\frac{1}{1-t}\right)
  +\,t\left(\frac{3}{2}\,t-1\right)\right]\hbox{\quad,\quad}t\;\equiv\; z/R_1
\end{equation}
Dividing Eq.\,(\ref{eqn:x2zweakOveras}) by Eq.\,(\ref{eqn:x2R1weakOveras}) and ignoring the difference between $f_\mathrm{dM,\,z/2}$ and $f_\mathrm{dM,\,R_1/2}$ we find 
\begin{equation}\label{eqn:PKuniversal}
\frac{\sigma_x(z)}{\sigma_x(R_1)}\;=\;\left[2\,(1-t)^2\ln\left(\frac{1}{1-t}\right)
  +\,3\,t^2-2\,t\right]^{1/2}\hbox{\quad,\quad}t\;\equiv\; z/R_1
\end{equation}
This is PK's second result: pencil beam spreading is universal if $\sigma$ is normalized to its maximum value and depth is normalized to its maximum value (range). We shall find this remains true for any heavy ion.

\begin{table}[p]
\begin{center}
\tabcolsep=4.5pt
\begin{tabular}{lrrrrrrrrrr}
\multicolumn{1}{c}{material}&           
\multicolumn{1}{c}{$\rho$}&           
\multicolumn{1}{c}{$\rho X_0$}&           
\multicolumn{1}{c}{$\rho X_S$}&           
\multicolumn{1}{c}{$k$}&           
\multicolumn{1}{c}{$a$}&           
\multicolumn{1}{c}{rms}&           
\multicolumn{1}{c}{$A$}&           
\multicolumn{1}{c}{rms}&           
\multicolumn{1}{c}{rms}&           
\multicolumn{1}{c}{$\sigma_x/R_1$}\\           

\multicolumn{1}{c}{}&           
\multicolumn{1}{c}{g/cm$^3$}&           
\multicolumn{1}{c}{g/cm$^2$}&           
\multicolumn{1}{c}{g/cm$^2$}&           
\multicolumn{1}{c}{\%}&           
\multicolumn{1}{c}{}&           
\multicolumn{1}{c}{\%}&           
\multicolumn{1}{c}{}&           
\multicolumn{1}{c}{\%}&           
\multicolumn{1}{c}{\%}&           
\multicolumn{1}{c}{\%}\\~\\

\multicolumn{1}{c}{1}&           
\multicolumn{1}{c}{2}&           
\multicolumn{1}{c}{3}&           
\multicolumn{1}{c}{4}&           
\multicolumn{1}{c}{5}&           
\multicolumn{1}{c}{6}&           
\multicolumn{1}{c}{7}&           
\multicolumn{1}{c}{8}&           
\multicolumn{1}{c}{9}&           
\multicolumn{1}{c}{10}&           
\multicolumn{1}{c}{11}\\~\\

HELIUM         &(0.166)&   94.32&  155.55&    6.67&   3983&   0.02&   4859&   3.85&   0.10&   1.67\\
HYDROGEN       &(0.084)&   61.28&  145.76&    6.11&   9221&   0.01&  10552&   3.52&   0.02&   1.16\\
BERYLLIUM      &  1.848&   65.19&   92.59&    6.98&   3285&   0.04&   4093&   4.03&   0.14&   1.72\\
POLYETHYLENE   &  0.940&   44.64&   61.79&    6.97&   4362&   0.02&   5333&   4.02&   0.12&   1.89\\
PILOTB         &  1.020&   44.23&   59.43&    7.01&   4097&   0.03&   5036&   4.05&   0.14&   1.98\\
POLYSTYRENE    &  1.060&   43.72&   59.15&    7.11&   3979&   0.03&   4914&   4.11&   0.15&   2.01\\
CARBLD         &  1.800&   42.70&   56.34&    7.18&   3582&   0.05&   4463&   4.15&   0.18&   2.12\\
CARBON         &  1.900&   42.70&   56.34&    7.18&   3582&   0.05&   4463&   4.15&   0.18&   2.11\\
LEXAN          &  1.200&   41.46&   55.05&    7.18&   3853&   0.03&   4777&   4.15&   0.17&   2.10\\
NYLON          &  1.130&   41.84&   54.73&    7.06&   4113&   0.03&   5061&   4.08&   0.14&   2.05\\
LUCITE         &  1.190&   40.59&   53.76&    7.18&   3943&   0.03&   4881&   4.15&   0.16&   2.10\\
KAPTON         &  1.420&   40.56&   53.21&    7.26&   3688&   0.04&   4596&   4.20&   0.19&   2.16\\
WATER          &  1.000&   36.08&   46.88&    7.29&   4007&   0.04&   4970&   4.21&   0.18&   2.25\\
AIR            &(1.205)&   36.66&   46.76&    7.32&   3539&   0.04&   4431&   4.23&   0.28&   3.05\\
TEFLON         &  2.200&   34.84&   43.87&    7.49&   3305&   0.05&   4178&   4.34&   0.26&   2.46\\
QUARTZ         &  2.640&   27.05&   32.94&    7.89&   3247&   0.06&   4156&   4.57&   0.35&   2.82\\
MAGNESIUM      &  1.740&   25.03&   30.17&    7.98&   3161&   0.06&   4065&   4.63&   0.40&   3.03\\
ALUMINUM       &  2.699&   24.01&   28.75&    8.12&   3007&   0.07&   3897&   4.70&   0.45&   3.11\\
SILICON        &  2.330&   21.82&   25.93&    8.11&   3099&   0.07&   4006&   4.70&   0.45&   3.25\\
IRON           &  7.874&   13.84&   15.82&    8.85&   2576&   0.13&   3453&   5.15&   0.86&   4.24\\
COPPER         &  8.960&   12.86&   14.62&    9.13&   2446&   0.13&   3321&   5.32&   1.01&   4.45\\
NICKEL         &  8.900&   12.68&   14.42&    9.15&   2600&   0.11&   3513&   5.32&   0.94&   4.36\\
ZINC           &  7.133&   12.43&   14.09&    9.27&   2457&   0.12&   3348&   5.40&   1.06&   4.56\\
BRASS          &  8.489&   12.30&   13.89&    9.51&   2403&   0.03&   3303&   5.46&   1.14&   4.58\\
MOLYBDENUM     & 10.220&    9.80&   10.91&    9.77&   2168&   0.15&   3033&   5.70&   1.54&   5.33\\
TIN            &  7.310&    8.82&    9.70&   10.13&   1995&   0.16&   2843&   5.92&   1.96&   5.91\\
GADOLINIUM     &  7.900&    7.48&    8.04&   10.24&   1839&   0.15&   2650&   5.97&   2.44&   6.69\\
TANTALUM       & 16.600&    6.82&    7.21&   11.18&   1717&   0.20&   2566&   6.56&   3.04&   6.82\\
MERCURY        & 13.550&    6.44&    6.72&   11.23&   1628&   0.21&   2450&   6.58&   3.44&   7.31\\
LEAD           & 11.350&    6.37&    6.63&   11.18&   1607&   0.17&   2418&   6.54&   3.54&   7.49\\
URANIUM        & 18.950&    6.00&    6.12&   11.37&   1525&   0.19&   2321&   6.65&   3.91&   7.71\\

\end{tabular}
\caption{Quantities related to the calculation of $\sigma_\mathrm{x,\,R_1}/R_1$ in various materials. Eq.\,\ref{eqn:x2R1Overas} is evaluated at five equally spaced ranges $R_1$, the largest corresponding to 250\,MeV incident, and a straight line passing through the origin is fitted cf. PK Figure\,3. Col.\,11 is the slope of that line and col.\,10 is the rms scatter of the five computed points about the line, a measure of the proportionality of $\sigma_\mathrm{x,\,R_1}$ to $R_1$. Col.\,2 is the density (parentheses: mg/cm$^3$), col.\,3 is the mass radiation length and col.\,4 is the mass scattering length. Cols.\,5,\,6 are the {\O}ver{\aa}s parameters from a two-point fit at $R_1/3$ and $2R_1/3$ and col.\,7 is the rms scatter (at the five ranges) about that fit, a measure of the goodness of the approximation. Col.\,8 is the `weak {\O}ver{\aa}s' parameter fitted at $R_1/2$ and col.\,9, the rms scatter about that fit.\label{tbl:Overas}}
\end{center}
\end{table}

Table\,\ref{tbl:Overas} lists parameters useful in evaluating the foregoing formulas for protons in 31 materials as well as the final constant of proportionality between $\sigma$ at end-of-range and range (see caption for details). From Table\,\ref{tbl:Overas}
\begin{eqnarray*}
\hbox{protons in water:\quad}\sigma_\mathrm{x,\,R_1}&=&2.25\%\times R_1\hbox{\qquad (PK Eq.\,(12c):\ \ 2.17\%)}\\
\hbox{protons in aluminum:\quad}\sigma_\mathrm{x,\,R_1}&=&3.11\%\times R_1\hbox{\qquad (PK Eq.\,(12d):\ \ 3.18\%)}
\end{eqnarray*}
so our rederived numerical values are very near PK's (taking into account, of course, their use of $\sigma_r=\sqrt{2}\times\sigma_x$ rather than $\sigma_x$).

\subsection{Generalization to Heavy Ions}
The results just obtained can be generalized to heavy ions. First we need to establish the range-energy relation for heavy ions and its consequence for $pv$. From stopping power theory \cite{icru49} the mean projected range of any fast charged particle is
\begin{equation}\label{eqn:runiver}
R\;=\;\frac{mc^2}{z^2}\frac{1}{c_B}\frac{A}{Z}\int_0^\tau\frac{\beta^2}{L(\beta)}\,d\tau'
\end{equation}
$mc^2$ is the particle's rest energy and $z$ its charge number. $\tau\equiv E/mc^2$ is the particle's reduced kinetic energy.\,\footnote{~The lower limit of the integral is actually some very small reduced energy, a cutoff parameter whose exact value does not matter.} $c_B$ is a universal constant while $Z$ and $A$ or (for compounds and mixtures) some combination of constituent $Z$s and $A$s, are atomic constants of the stopping material. $L(\beta)$ is the `stopping number', a function which, to a very good approximation, depends only on speed \cite{icru49}. From Eq.\,(\ref{eqn:betasq}) an ion's speed depends only on its $\tau=E/mc^2$. Therefore, Eq.\,(\ref{eqn:runiver}) implies
\[1\;=\;\frac{R_\mathrm{I}}{R_p}\;=\;\frac{(mc^2/z^2)_\mathrm{I}}{(mc^2/z^2)_p}\]
as long as
\[\tau_p\;=\;\tau_\mathrm{I}\hbox{\qquad or\qquad}E_p\;=\;\frac{m_p}{m_\mathrm{I}}\,E_\mathrm{I}\]
Introducing $a_\mathrm{I}\equiv m_\mathrm{I}/m_p$ and noting that $z_p=1$ we find for any ion
\begin{equation}\label{eqn:Ri}
R_\mathrm{I}(E_\mathrm{I})\;=\;\frac{a_\mathrm{I}}{z_\mathrm{I}^2}\;R_p\Big(\frac{E_\mathrm{I}}{a_\mathrm{I}}\Big)
\end{equation}
where $R_p$ is proton range as a function of energy.\,\footnote{~Of course $R_p(E)$ and $E_p(R)$ also depend on the stopping material.} From this, if $E_p$ is proton energy as a function of range,
\begin{equation}\label{eqn:Ei}
E_\mathrm{I}(R_\mathrm{I})\;=\;a_\mathrm{I}\,E_p\Big(\frac{z_\mathrm{I}^2}{a_\mathrm{I}}R\Big)
\end{equation}
Given a proton range-energy function and its inverse (which must now extend to very large ranges because of $z_\mathrm{I}^2/a_\mathrm{I}$) we can now solve any range-energy problem for any ion.\footnote{~Our standard proton functions Range(energy,\;material) and Energy(range,\;material) assume kinetic energy in MeV and range in g/cm$^2$, that is, mass range $\rho R$, despite consistently calling it `range'.}

The {\O}ver{\aa}s approximations also generalize, as shown by Kanematsu \cite{kanematsu08}. The weak version Eq.\,(\ref{eqn:weakOveras}) reads, for protons,
\begin{equation}\label{eqn:pv2p}
(pv)_p^2\;=\;A_p\,(\rho R)_p
\end{equation}
Suppose a proton and an ion have the same speed $v$. From special relativity 
\[pc\;=\;mc^2/\sqrt{1-\beta^2}\] 
so momentum is proportional to mass and therefore
\[(pv)_p^2\;=\;(pv)_\mathrm{I}^2/a_\mathrm{I}^2\]
As already shown
\[(\rho R)_p\;=\;(z_\mathrm{I}^2/a_\mathrm{I})\,(\rho R)_\mathrm{I}\]
Putting it all together we get the weak {\O}ver{\aa}s approximation for an ion in terms of the `{\O}ver{\aa}s constant' of a proton
\begin{equation}\label{eqn:pv2x}
(pv)_\mathrm{I}^2\;=\;(a_\mathrm{I}\,z_\mathrm{I}^2)\,(\rho R)_\mathrm{I}\,A_p
\end{equation}
Kanematsu \cite{kanematsu08} gives the analogous `strong' approximation for water, also showing that it is much more accurate than the usual power law $R=aE^b$. We will not need it here, however.

To complete the generalization to heavy ions we need to consider the effect on scattering power. On elementary considerations (the Coulomb force) we must attach a factor $z_\mathrm{I}$ to every occurrence of $E_s$ \cite{rossiBook}, which we have already done. Kanematsu's scattering power $T_\mathrm{dH}$ \cite{kanematsu08} requires no further modification since its nonlocal correction factor depends only on the path integral of $(1/X_0)$. 

We wish, however, to use our $T_\mathrm{dM}$ (Eqs.\,\ref{eqn:TdM},\,\ref{eqn:fdM}) whose correction factor $f_\mathrm{dM}$ depends logarithmically on the diminution of $pv$. $f_\mathrm{dM}$ as it stands (Eq.\,\ref{eqn:fdM}) will not work since it was optimized for protons and characteristic $(pv)$s for ions of the same range are much larger, as we have just shown. Rather than opening Pandora's box and re-optimizing $f_\mathrm{dM}$ for each ion let us apply an appropriate reduction factor $f_\mathrm{I}$ to both $p_1v_1$ and $pv$ letting
\[f_\mathrm{dM}(p_1v_1,\,pv)\rightarrow f_\mathrm{dM}(f_\mathrm{I}\times(p_1v_1),\,f_\mathrm{I}\times(pv))\]
The form of $f_\mathrm{I}$ is suggested by Eqs.\,(\ref{eqn:pv2p},\,\ref{eqn:pv2x}) which, for protons and ions of the same range, give
\begin{equation}\label{eqn:pvp}
(pv)_p\;=\;(pv)_\mathrm{I}/(z_\mathrm{I}\sqrt{a_\mathrm{I}})
\end{equation}
A somewhat better approximation is
\begin{equation}\label{eqn:fdMi}
f_\mathrm{I}\;=\;1/(z_\mathrm{I}\sqrt{a_\mathrm{I}})^{1.018}
\end{equation}
Figure\,\ref{fig:PVratio} shows it is good to approximately $\pm2\%$ for materials, ions and ranges of clinical interest. That is good enough, because $f_\mathrm{dM}$ only depends logarithmically on $pv$. For protons, $f_\mathrm{I}=1$. 

Also involved in the derivation of $T_\mathrm{dM}$ \cite{scatPower2010} was a condition on the characteristic large-angle (nuclear size) cutoff of Rutherford scattering. That lead to the simplification of $\chi_2/\chi_1$ and thus to the introduction of scattering length $X_S$.  As easily shown, $\chi_2<1$ is even better satisfied for heavy ions than for protons.

Putting everything together, the derivations of the previous section go through exactly as before and we find that Eq.\,(\ref{eqn:sigxoR1simple}) for the $R_1$ coefficient of rms beam spread and Eq.\,(\ref{eqn:PKuniversal}) for the universal beam spreading profile apply also to heavy ions. The influence of heavy ions in Eq.\,(\ref{eqn:sigxoR1simple}) is found entirely in $z_\mathrm{I}$ and the fact that $pv$ for heavy ions is much larger.

Combining Eqs.\,(\ref{eqn:pvp}) and (\ref{eqn:sigxoR1simple}) we also find, for ions and protons of the same range at any depth
\begin{equation}\label{eqn:sxi}
\sigma_{x,\,\mathrm{I}}\;=\;\sigma_{x,\,p}/\sqrt{a_\mathrm{I}}
\end{equation} 
Kanematsu \cite{kanematsuLD} obtains $z_\mathrm{I}^{-0.08}\,a_\mathrm{I}^{-0.46}$, nearly the same. The reason for the disappearance or near disappearance of $z$ is fundamental. Stopping power and scattering power both vary as $z^2$. The extra Coulomb scattering is offset by the increased speed needed to get the same range. 

\begin{figure}[p]
\centering\includegraphics[width=3.64in,height=3.5in]{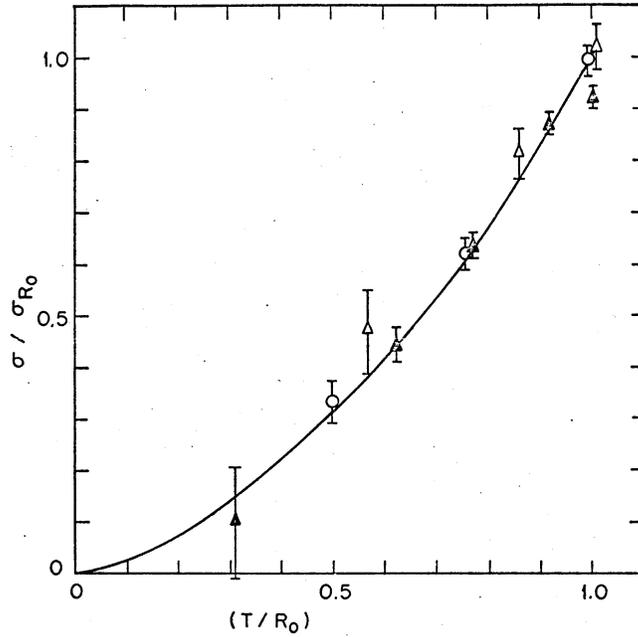} 
\caption{Facsimile of PK Figure\,17 whose caption reads ``Dimensionless plot of the standard deviation due to scattering versus depth of penetration $\ldots$ Open triangles are experimental results for 112\,MeV protons on aluminum; solid triangles for 158\,MeV protons on aluminum; open circles for 127\,MeV protons on water."\label{fig:pk17}}
\end{figure}
\begin{figure}[p]
\centering\includegraphics[width=4.50in,height=3.5in]{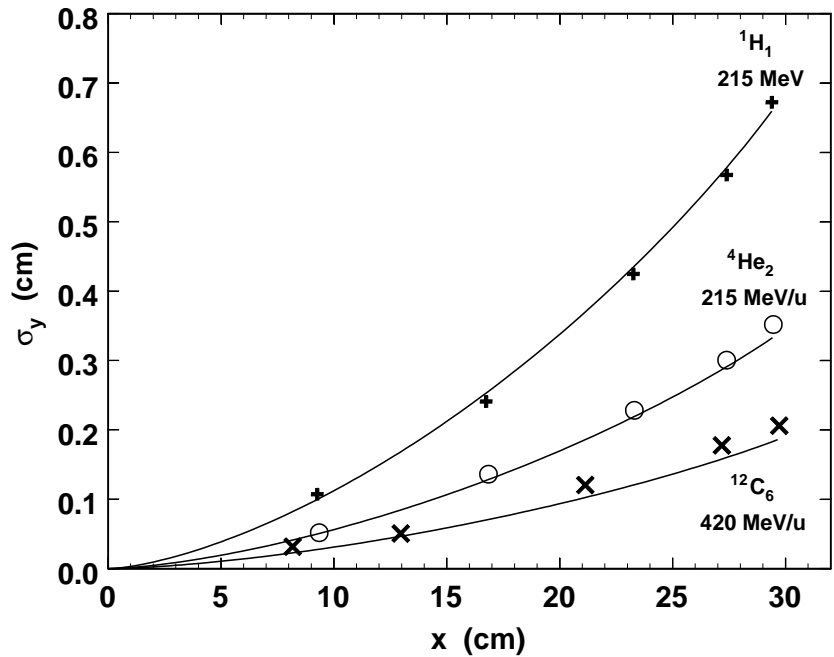} 
\caption{Comparison between `data from M. Phillips, LBL' (see text) and PK theory generalized to heavy ions (Eqs.\,\ref{eqn:sigxoR1simple},\,\ref{eqn:PKuniversal}) (lines). \label{fig:Phillips}}
\end{figure}

\subsection{Experimental Tests}
A number of transverse scans of heavy ion beams vs. depth are available. Unfortunately they tend to be motivated entirely by clinical applications (beam commissioning), thus dominated by ion-optical effects or beam contamination \cite{Schwaab2011} and not too useful as tests of the basic theory. After moderate effort we have found only the following three.

\subsubsection{Preston and Koehler}
The best (perhaps only) proton data remain those of PK \cite{preston}. Protons of approximately 158\,MeV from the Harvard Cyclotron, degraded to lower energy as required and collimated to a 2\,mm diameter beam, fell on a target of adjustable thickness. The beam profile exiting the target was measured using a very small remotely driven diode dosimeter \cite{AMKdiode} in conjunction with a plane parallel ion chamber beam monitor. The observed Gaussian distributions were fitted graphically to extract $\sigma$. In-air $\sigma$\,s at the same depths were subtracted in quadrature to correct for beam and detector size.

PK's Figure\,17 (our Figure\,\ref{fig:pk17}) shows their universal curve along with data from two energies entering aluminum and one energy entering water. The data agree with theory (our Eqs.\,\ref{eqn:sigxoR1simple} and \ref{eqn:PKuniversal}) within the experimental error of a few percent.

\subsubsection{Phillips}
This is a curious tale. A 1990 paper by Kraft \cite{Kraft1990} refers to `data from M. Phillips, LBL' in a figure caption, with no further citation. Some of those data (for $^1$H$_1$, $^4$He$_2$ and $^{12}$C$_6$) were picked up by Hollmark et al. \cite{hollmark} who cited Kraft, and corrected for the size of the incident beam.\,\footnote{~That correction, obtained by extrapolation, corresponds to an unrealistically small beam diameter of 1.2\,mm. It would, in any case, almost certainly have been done already by the original authors.} Finally, Kanematsu \cite{kanematsu08} used the data referring to `Phillips' measurements' and citing Hollmark. The data also show up in a Master of Science thesis by M. Granlund (2001) and perhaps elsewhere.

LBL issued numerous technical notes at the time and it seemed unlikely that an experiment so difficult and fundamental would not have been written up or published. However, a library search turned up nothing. Aided by Prof. Kraft we located Prof. Phillips who, unfortunately, had discarded his documents from that period and remembered little of the episode: 
\begin{quote}
`$\ldots$ I suspect that the data that Gerhard Kraft originally published was a set of measurements I collated from the LBL particle therapy group (J. Lyman, W. Chu, T. Renner, B. Ludewigt) or actually just obtained from the treatment planning code data files (which of course would be data from the same group) $\ldots$'  
\end{quote}
He did, however, make us aware of the important LBL measurement to come next.

The purpose of Kraft's paper was to promote heavy ion therapy and any reasonably correct data showing that heavy ions scatter less would have served. All things considered, we doubt that the data represent a measurement at all. However, in keeping with tradition, we compare the theory just derived with M. Phillips' data, read from Hollmark's graph, in Figure\,\ref{fig:Phillips}. We have not corrected for incident beam size. The agreement is quite good.

\subsubsection{Wong et al.}
This \cite{Wong1990} is not a measurement of beam spreading but nevertheless extremely valuable as it confirms our understanding of MCS of heavy ions in an extreme case, $^{238}$U$_{92}$ scattering in Cu.

Partly stripped U ions accelerated to 650\,MeV/$a_\mathrm{I}$ passed through a four element position-sensitive Si detector array, a 0.27\,cm Cu target, and a second array. One assumes the U stripped completely on entering the Si. Each array made two $x$ and two $y$ measurements to establish incoming and outgoing track segments so the projected $x$ and $y$ deflections by the Cu were measured independently event-by-event. The final detector was thicker to provide a good $\Delta E$ measurement for background rejection.

The analysis culminated in a Monte Carlo simulation using, in turn, three different MCS models namely $\theta_\mathrm{FR}$ and $\theta_\mathrm{Highland}$ (in our terminology \cite{scatPower2010}) and the full Moli\`ere treatment of projected angle \cite{moliere2}. The MCS physics is thus somewhat buried, and the difficulty \cite{kanematsu08,scatPower2010} of incorporating Highland's equation in a Monte Carlo  was perhaps not fully appreciated. The authors concluded that Highland fit the final projected angle distribution best (it fits exactly), Moli\`ere next (too broad) and $\theta_\mathrm{FR}$ worst (even broader). Nevertheless
\begin{quote}
`$\ldots$\,all of the theories predicted a Gaussian distribution in satisfactory agreement with experiment, although the modified Fermi theory reproduced the data best\,$\ldots$'
\end{quote} 
as close as the authors come to giving an experimental error.

Fortunately a much simpler analysis is possible using the information provided. Figure\;6 gives target-in and target-out projected angular distributions for $x$ and $y$. Scattering in Cu can be obtained by subtraction in quadrature, giving almost equal $x$ and $y$ angles which can be averaged and compared directly with various theoretical predictions using the mixed slab procedures and notation of \cite{scatPower2010}. In order of agreement with experiment (last column), with Kanematsu's `differential Highland' scattering power \cite{kanematsu08} leading the pack:
\begin{table}[h]
\tabcolsep=10pt
\begin{center}
\begin{tabular}{lrr}
\multicolumn{1}{l}{description}&           
\multicolumn{1}{c}{mrad}&           
\multicolumn{1}{c}{\%\rule[-6pt]{0pt}{6pt}}\\
$\theta_\mathrm{measured}$&2.607&\rule[-6pt]{0pt}{6pt}\\
$\theta_\mathrm{dH}$&2.625&0.66\\
$\theta_\mathrm{Highland}$&2.642&1.35\\
$\theta_\mathrm{dM}$&2.683&2.88\\
$\theta_\mathrm{Hanson}$&2.755&5.67\\
\end{tabular}
\end{center}
\end{table}

Of course, stopping powers or range-energy relations enter any such analysis or Monte Carlo. Those used by the authors were not directly described in the paper, and a citation proved unhelpful. Be that as it may, we used the heavy-ion formulas of the preceding section, the ICRU49 proton range-energy tables \cite{icru49}, guessed at those Si detector thicknesses not given, combined each detector array and ignored air obtaining finally

\begin{table}[h]
\tabcolsep=8pt
\begin{center}
\begin{tabular}{ccccc}
\multicolumn{1}{c}{slab}&           
\multicolumn{1}{c}{material}&           
\multicolumn{1}{c}{cm}&           
\multicolumn{1}{c}{MeV/a}&           
\multicolumn{1}{c}{res cm Cu}\rule[-6pt]{0pt}{6pt}\\
1&Si&0.3654&650.0&0.799\\
2&Cu&0.2700&587.2&0.686\\
3&Si&0.5324&424.7&0.416\\
4&-&-&310.5&0.252\\
\end{tabular}
\end{center}
\end{table}
Col.\,3 is the slab thickness, Col.\,4 is $E$ entering the slab and Col.\,4 is the residual range in Cu entering the slab. The only range-related benchmark in the paper is somewhat incidental, citing the maximum Cu that might have been used:
\begin{quote} 
`$\ldots$\ corresponding to uranium stopping in the final Y4 detector; it corresponds to a calculated copper target thickness of 3.5\,mm\ $\ldots$'
\end{quote}
whereas we obtain 5.15\,mm. We do not understand the discrepancy, but the good agreement obtained by Wong et al. suggest that their range-energy treatment was in fact correct.

Our ranking of $\theta_\mathrm{Highland}$ and $\theta_\mathrm{Hanson}$ agrees with the authors' Monte-Carlo based conclusions. The amazingly good agreement of all four MCS theories with measurement must be fortuitous considering the likely experimental error. Still, the results of this experimental {\em tour de force} (2.6\,mrad corresponds to 2.6\,mm in 1\,m\,!) strongly reinforce our confidence in the MCS theory of heavy ions, as Wong et al. concluded.

\subsection{Summary}
\begin{quote}
We rederived PK's results \cite{preston} in Fermi-Eyges language.

PK find the equivalent of Eq.\,(\ref{eqn:A2}) by reasoning that the rms transverse size of the beam at an arbitrary measuring plane is the quadratic sum of displacements, projected onto that plane, from infinitesimal elements of the stack. All their results follow from this single fragment of Fermi-Eyges theory.

If the single scattering correction in the integral of Eq.\,(\ref{eqn:A2}) is replaced by its effective value and $pv$ is replaced by the {\O}ver{\aa}s approximation then $<x^2_\mathrm{max}>$ (at $z=R_1$) can be found in closed form (Eq.\,\ref{eqn:x2R1Overas}). $<x^2_\mathrm{max}>$  is very nearly proportional to $R_1^2$.

If, instead, $pv$ is replaced by the weak {\O}ver{\aa}s approximation, then $<x^2_\mathrm{max}>$ is exactly proportional to $R_1^2$ and {\em the end-point rms transverse beam size is proportional to range} (PK's first result). The constant of proportionality is given by Eq.\,(\ref{eqn:sigxoR1simple}).

In the weak {\O}ver{\aa}s approximation the integral can also be evaluated at arbitrary depth leading (Eq.\,\ref{eqn:PKuniversal}) to a {\em universal form for beam spreading in terms of normalized depth} (PK's second result). It holds for protons of any incident energy traversing any material.

The PK results can be generalized to arbitrary ions of mass/(proton mass) $a_\mathrm{I}$ and charge number $z_\mathrm{I}$. Because stopping power depends only on these and speed, the range of an ion\,, given its kinetic energy, can be scaled from the proton range-energy relation (Eq.\,\ref{eqn:Ri}) and the energy of an ion, given its range, from the corresponding proton relation (Eq.\,\ref{eqn:Ei}). The weak {\O}ver{\aa}s approximation scales similarly (Eq.\,\ref{eqn:pv2x}).

The scattering power for heavy ions depends on the single scattering correction. If, as in $T_\mathrm{dM}$, it is a function of $pv$ it must be adjusted for the fact that $pv$ for ions is much greater than for protons of the same range. Eq.\,\ref{eqn:fdMi} gives the appropriate factor to reduce $pv$ to `proton scale' and thus restore the validity of $T_\mathrm{dM}$\;.  

With all this done, the PK derivation goes through as before, extending their results to heavy ions. Beam spreading at any depth is less for a heavy ion than for a proton of the same range by $1/\sqrt{a_\mathrm{I}}$ (Eq.\,\ref{eqn:sxi}).

There are few experimental tests of beam spreading in matter. PK validated their own theory with measurements \cite{preston} at several proton energies in aluminum and water. Data from Phillips at LBL have been widely cited but may or may not come from experiment. Wong et al. \cite{Wong1990} is not a beam spreading measurement but nicely confirms multiple Coulomb scattering theory in an extreme case, uranium scattering in copper.
\end{quote}

\section{Postscript and Acknowledgements}
In radiotherapy, Fermi-Eyges theory was first used with electron beams. It is rarely mentioned in early proton papers even where it might be appropriate. That is paradoxical because protons are much better Fermi-Eyges particles. The small angle approximation always applies, and the relation between energy and depth holds up longer because range straggling is much smaller. However, protons came to radiotherapy from the high-energy world where (also paradoxically) Fermi-Eyges theory is not so well known. 

Thus our own education in the theory came late, and we thank Sairos Safai, Uwe Schneider and above all Nobuyuki Kanematsu for contributing to it. We thank LBL librarians Sarah Cruz and Michael Golden who searched for the elusive Phillips data. We are as always greatly indebted to Harvard University, represented by the Laboratory for Particle Physics and Cosmology, for an office, Internet connection, and the other resources that flow from a courtesy appointment to the Physics Department.

Finally, we thank Bill Preston, Andy Koehler and Miles Wagner. Their own work advanced the scientific and technical development of the field, but more than that, as successive Directors of the Harvard Cyclotron Lab (1948\,-\,2002) they placed a high priority on research, development and dissemination of knowledge in addition to patient throughput. The author, and ultimately this work, benefited greatly from the time and space that provided.

\appendix
\section{Analytical Geometry of the Ellipse}\label{sec:ellipse}
Following Born and Wolf \cite{BornWolf} we derive relations between parameters of the ellipse, centered on the origin, in several forms: the parametric form in the usual $x,y$ frame, the parametric form in a $u,v$ frame tilted to coincide with the principal axes, and the quadratic form in both frames. We also prove a formula for the area enclosed by a tilted ellipse.

\begin{figure}[h]
\centering\includegraphics[width=4.32in,height=3.5in]{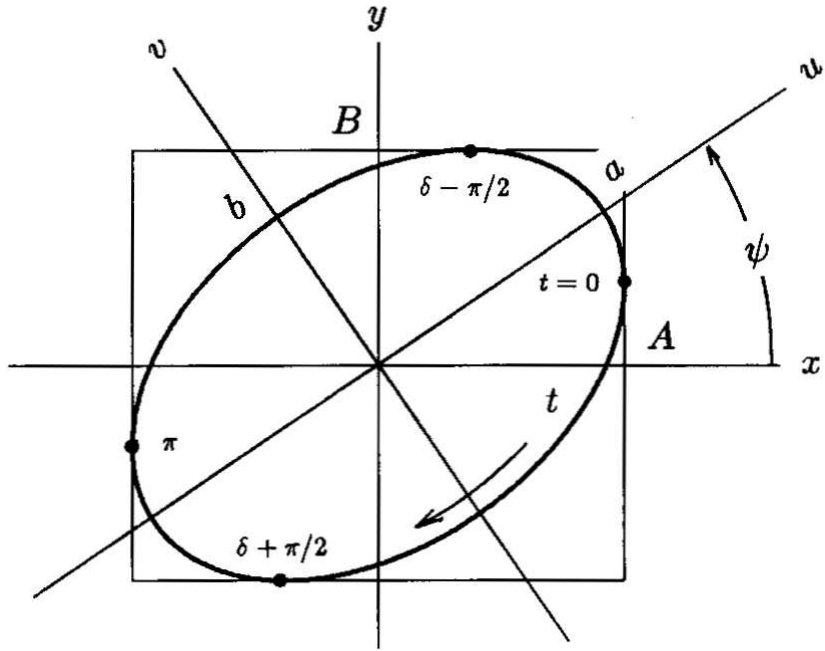} 
\caption{Ellipse geometry.\label{fig:ellipse}}
\end{figure}

\subsection{Tilted Ellipse}
The parametric form is
\begin{eqnarray}
x\;&=&\;A\cos t\label{eqn:x}\\
y\;&=&\;-B\sin(t-\delta)\label{eqn:y}
\end{eqnarray}
By inspection this defines a closed curve depending on three parameters $A,B,\delta$~, touching the boun\-ding box $2A\times2B$ at the points labeled in Figure\;\ref{fig:ellipse}. If $\delta=0$, squaring and adding yields the quadratic form
\begin{equation}\Bigl(\frac{x}{A}\Bigr)^2\;+\;\Bigl(\frac{y}{B}\Bigr)^2\;=\;1\end{equation}
To express the more general tilted ellipse similarly we first rewrite Eqs.\,(\ref{eqn:x},\;\ref{eqn:y}) as
\begin{eqnarray}
\frac{x}{A}\;&=&\;\cos t\label{eqn:xoA}\\
\frac{y}{B}\;&=&\;-\sin t\cos \delta\;+\;\cos t\sin \delta\label{eqn:yoB}
\end{eqnarray}
Taking [Eq.\,(\ref{eqn:xoA})\,$\times\cos\delta$] and [Eq.\,(\ref{eqn:xoA})\,$\times\sin\delta$ - Eq.\,(\ref{eqn:yoB})] and simplifying we obtain
\begin{eqnarray}
\frac{x}{A}\;\cos\,\delta&=&\cos\,t\;\cos\,\delta\nonumber\\
\frac{x}{A}\;\sin\,\delta\;-\;\frac{y}{B}&=&\sin\,t\;\cos\,\delta\nonumber
\end{eqnarray}
Squaring, adding and simplifying yields
\begin{equation}
\Bigl(\frac{x}{A}\Bigr)^2-\;2\sin\;\delta\Bigl(\frac{x}{A}\Bigr)\Bigl(\frac{y}{B}\Bigr)\;+\;\Bigl(\frac{y}{B}\Bigr)^2\;
  =\;\cos^2\delta\label{eqn:conic}
\end{equation}
the equation of a conic section, an ellipse or circle because the discriminant is negative: 
\[4\sin^2\delta/(A^2B^2)-\,4/(A^2B^2)=-\,(2\cos\delta/AB)^2\le0\]

\subsection{Transformation to Principal Frame}
Let us now find the parameters of the same ellipse in the frame $u,v$ tilted with respect to $x,y$ by an angle $\psi$ such that $u$ and $v$ coincide with the principal axes of the ellipse (see Figure \ref{fig:ellipse}). The transformation from $x,y$ to $u,v$ is a counterclockwise rotation through $\psi$
\begin{eqnarray}
u&=&x\cos\psi+y\sin\psi\label{eqn:urot}\\
v&=&-\,x\sin\psi+y\cos\psi\label{eqn:vrot}
\end{eqnarray}
and the ellipse has the new and simpler form
\begin{eqnarray}
u&=&a\cos(t-t_0)\label{eqn:u}\\
v&=&-\,b\sin(t-t_0)\label{eqn:v}
\end{eqnarray}
We assume $a$ is the semimajor axis, $a\ge b$. $t_0$ is not really an ellipse parameter since changing it does not change the shape or orientation of the ellipse. It is simply a shift in the origin of the time-like parameter $t$ to ensure that Eqs.\,(\ref{eqn:x},\;\ref{eqn:y}) and Eqs.\,(\ref{eqn:u},\;\ref{eqn:v}) are capable of describing the same situation even with respect to the initial location on the ellipse. Since these equations are supposed to describe the same ellipse we have, on using Eqs.\,(\ref{eqn:urot},\,\ref{eqn:vrot}),
\begin{eqnarray}
a\cos(t-t_0)&=&A\cos t\cos\psi-B\sin(t-\delta)\sin\psi\nonumber\\
-\,b\sin(t-t_0)&=&-\,A\cos t\sin\psi-B\sin(t-\delta)\cos\psi\nonumber
\end{eqnarray}
Expanding, we obtain
\begin{eqnarray}
a\;(\cos t\cos t_0+\sin t\sin t_0)&=&A\cos t\cos\psi-B\sin\psi(\sin t\cos\delta-\cos t\sin\delta)\nonumber\\
b\;(\sin t\cos t_0-\cos t\sin t_0)&=&A\cos t\sin\psi+B\cos\psi(\sin t\cos\delta-\cos t\sin\delta)\nonumber
\end{eqnarray}
If these are to hold identically the coefficients of $\sin t$ and $\cos t$ must be equal:
\begin{eqnarray}
a\cos t_0&=&A\cos\psi+B\sin\psi\sin\delta\label{eqn:acos}\\
a\sin t_0&=&-\,B\sin\psi\cos\delta\label{eqn:asin}\\
b\sin t_0&=&-\,A\sin\psi+B\cos\psi\sin\delta\label{eqn:bsin}\\
b\cos t_0&=&B\cos\psi\cos\delta\label{eqn:bcos}
\end{eqnarray} 
These four equations contain all possible relations between $a,b,\psi$ and $A,B,\delta$. One such is obtained by squaring each, adding and simplifying:
\begin{equation}
a^2+b^2=A^2+B^2\label{eqn:asq}
\end{equation}
Another is obtained by taking Eq.\,(\ref{eqn:acos})$\times$Eq.\,(\ref{eqn:bcos})\,+\,Eq.\,(\ref{eqn:asin})$\times$Eq.\,(\ref{eqn:bsin}) and simplifying:
\begin{equation}
a\,b=A\,B\cos\delta\label{eqn:ab}
\end{equation}
Finally, equating Eq.\,(\ref{eqn:bsin})/Eq.\,(\ref{eqn:asin}) to Eq.\,(\ref{eqn:bcos})/Eq.\,(\ref{eqn:acos}), cross multiplying and simplifying we obtain
\begin{equation}
2AB\sin\delta\cos2\,\psi=(A^2-B^2)\sin2\,\psi\label{eqn:2AB}
\end{equation}
If we define an auxiliary angle $\alpha$ (where $0\le\alpha\le\pi/2$) as
\begin{equation}
\tan\alpha\equiv B/A\label{eqn:alpha}
\end{equation}
we can rewrite Eq.\,(\ref{eqn:2AB}) in the form
\begin{equation}
\tan2\,\psi=\tan2\,\alpha\sin\delta\label{eqn:tan2psi}
\end{equation}
which gives the tilt angle $\psi$ in terms of $A,B,\delta$. Dividing Eq.\,(\ref{eqn:ab}) by Eq.\,(\ref{eqn:asq}) and using the definition Eq.\,(\ref{eqn:alpha}) we obtain
\[
\frac{2ab}{a^2+b^2}=\frac{2AB}{A^2+B^2}\cos\delta=\sin2\,\alpha\cos\delta
\]
Defining a second auxiliary angle $\chi$ (where $0\le\chi\le\pi/4$) by
\begin{equation}
\tan\chi\equiv b/a\label{eqn:chi}
\end{equation} 
we can rewrite this as
\begin{equation}
\sin2\,\chi=\sin2\,\alpha\;\cos\delta\label{eqn:sin2chi}
\end{equation}

\subsection{Summary and Explicit Procedures}
Eqs.\,(\ref{eqn:x}, \ref{eqn:y}) and Eqs.\,(\ref{eqn:u}, \ref{eqn:v}) are the parametric form in the `lab' and `principal' frames respectively. Eqs.\,(\ref{eqn:alpha}) and (\ref{eqn:chi}) define useful auxiliary angles $\alpha$ and $\chi$, basically the aspect ratios in the two frames. Finally Eqs.\,(\ref{eqn:asq}), (\ref{eqn:tan2psi}) and (\ref{eqn:sin2chi}) relate $a,b,\psi$ to $A,B,\delta$ using the auxiliary angles. Given $A,B,\delta$ one may find $a,b,\psi$ using the procedure
\begin{eqnarray}
\alpha&=&\arctan B/A\nonumber\\
\chi&=&.5\arcsin\,(\sin2\,\alpha\cos\delta)\nonumber\\
\psi&=&.5\arctan\,(\tan2\,\alpha\sin\delta)\nonumber\\
a&=&\biggl(\frac{A^2+B^2}{1+\tan^2\chi}\biggr)^{1/2}\nonumber\\
b&=&a\tan\chi\nonumber
\end{eqnarray}
Given $a,b,\psi$ one may find $A,B,\delta$ using 
\begin{eqnarray}
\chi&=&\arctan b/a\nonumber\\
\alpha&=&.5\arcsin\;\biggl(\frac{\tan^22\,\psi+\sin^2 2\,\chi}{1+tan^2 2\,\psi}\biggr)^{1/2}\nonumber\\
\delta&=&\arcsin\,(\tan2\,\psi/\tan2\,\alpha)\nonumber\\
A&=&\biggl(\frac{a^2+b^2}{1+\tan^2\alpha}\biggr)^{1/2}\nonumber\\
B&=&A\tan\alpha\nonumber
\end{eqnarray} 

\subsection{Area Enclosed by the Ellipse}
The area enclosed\footnote{~It is incorrect to say `area of an ellipse'. The ellipse itself is a closed curve, having no area.} by an erect ellipse of semiaxes $a,b$ is easily obtained by integration:
\begin{eqnarray}
\hbox{area}&=&4\,\int_0^ab(1-(x/a)^2)^{1/2}dx\;=\;4\,a\,b\int_0^1(1-u^2)^{1/2}du\nonumber\\
&=&4\,a\,b\int_0^{\pi/2}\cos^2\theta\;d\theta\;=\;\pi\,a\,b\label{eqn:area}
\end{eqnarray}
For a more general and less well known equation we note that, from Eq.\,(\ref{eqn:conic}), $B\cos\delta$ is the $y$ intercept of the ellipse and we already know that $A$ is the maximum value of $x$. Therefore from Eqs.\,(\ref{eqn:ab}) and (\ref{eqn:area}) we find
\begin{equation}\hbox{area}\;=\;\pi\;y_{int}\;x_{max}\;=\;\pi\;x_{int}\;y_{max}\label{eqn:areatilt}\end{equation}
the second version following from symmetry. Eq.\,(\ref{eqn:areatilt}) reduces to Eq.\,(\ref{eqn:area}) for an erect ellipse. It is remarkable that Eq.\,(\ref{eqn:areatilt}) seems to be known only to accelerator physicists. We were unable to find it on the Web, unlike Eq.\,(\ref{eqn:area}) which is everywhere.

\listoffigures

\bibliographystyle{unsrt}
\bibliography{/pctexv4/work/pbs/master}

\end{document}